# Kinetically controlling surface atom arrangements in thermally robust, amorphous high-entropy alloy nanoparticles by solvent selection

Varatharaja Nallathambi[1,2], Se-Ho Kim[2,3], Baptiste Gault[2,4], Sven Reichenberger[1], Dierk Raabe[2*], Stephan Barcikowski[1*]

[1] Technical Chemistry I and Center for Nanointegration Duisburg-Essen (CENIDE), University of Duisburg-Essen, 45141 Essen, Germany

[2] Max Planck Institute for Sustainable Materials, Max-Planck-Str.1, 40237 Düsseldorf, Germany

[3] Department of Materials Science & Engineering, Korea University, Seoul, 02841, Republic of Korea

[4] Department of Materials, Royal School of Mines, Imperial College London, London SW72AZ, United Kingdom

* corresponding authors: stephan.barcikowski@uni-due.de, d.raabe@mpie.de

## Abstract

The ability to tailor nanoscale surface atom arrangements through multi-elemental compositional control provides high-entropy nanoalloys with promising functional properties. Developing a fundamental understanding of nanoalloy formation mechanisms during synthesis is therefore essential for effectively engineering the surface composition and resulting functional properties. Using the Cantor alloy (CrMnFeCoNi) as a model system, we investigate how solvent selection during reactive, nanosecond-pulsed laser synthesis influences carbon doping and the resulting changes in nanoparticle morphology, structure, and composition. Supersaturated carbon incorporation, partitioned from the organic solvent molecules, produces amorphous nanoparticles with distinctive carbon shells, thermally stable up to 350 °C. We propose kinetically controlled particle formation mechanisms and rationalize the criticality of the time scales between the competing reactions of carbon doping, carbon shell formation, and coalescence of metallic fragments, ruling compositional and morphological characteristics. This work demonstrates effective solvent-driven surface-compositional control in amorphous high-entropy nanoalloys. It introduces a novel synthesis approach for tailoring surface atom arrangements through carbon incorporation via reactive, pulsed laser synthesis.

# Main

Nanomaterials with their unique size and shape effects are attractive for a wide range of applications including electronics[1], optoelectronics[2], catalysis[3–5], biomedicine[6], energy conversion and storage[7]. Doping or alloying nanoparticles with solute elements is a well-established strategy to improve their functional properties[8–10]. Extending this concept to high-entropy alloys (HEA) or compositionally complex alloys (CCA), with multiple principal elements mixed in near-equiatomic proportions that form a single-phase solid solution, is a new frontier in heterogeneous catalysis[11–15] and in developing more efficient nanomaterials for high-temperature applications[16]. The concept of HEAs originally focused on solid solutions stabilized by enhanced configurational entropy and now the term broadly includes compositionally complex materials with extended solid solution ranges, and not limited to those stabilized solely by entropy[12,17]. To maintain similarity with the literature on this topic, we proceed to use the term HEA in the present work.

In bulk, HEAs have been explored for their promising mechanical, magnetic, and high-temperature behavior, and the ability to adjust composition and atomic configurations in these materials opens a practically infinite chemical and structural space for designing multifunctional nanoalloys[13,18–20]. The complex multi-elemental surface atom arrangements can induce unprecedented electronic and optical responses that open opportunities for optimizing the functional performance, for example, in catalyzing specific chemical reactions[12,21].

Achieving rational and controllable synthesis with precise control over the microstructure, morphology, and composition of HEAs remains a significant challenge21–23, particularly for nanomaterials with a high surface-to-volume ratio needed for catalytic activities[17,22,23]. The vast compositional space available to design HEA catalysts also faces limitations because of miscibility gaps and the contribution of the positive enthalpy of mixing between the constitutive binary and ternary systems involved, leading to phase separation[24,25]. In addition, the random mixing of the elemental components in nanostructures is difficult to achieve as most approaches require high temperatures, potentially favoring demixing, reacting, or stabilizing agents that can alter the chemistry, or controlled environments that complicate upscaling and control of the synthesis.

Laser synthesis and processing of colloids (LSPC) is a well-established and robust synthesis technique with the unique advantage of producing surfactant-free, colloidally-stable nanoparticles (NPs) with exceptional compositional freedom and high productivity[26–30]. LSPC encompasses a variety of techniques, namely laser ablation, fragmentation, melting, and reduction in liquids, and offers vast flexibility in the type, form, and composition of the source material and the solvent medium used[31]. In addition, the rapid heating and fast cooling rates during NP generation help the formation of thermodynamically metastable nanostructures, kinetically hindering the equilibration process[32,33]. Recent studies reported the synthesis of HEA NPs using laser ablation in liquids[21,34–38] as well as laser reduction of metal precursor mixtures[39]. The solvent used during the synthesis influences the nature of the reactive species formed from molecular decomposition during pulsed laser irradiation[40,41]. In organic solvents, the reactive carbon species released *in situ*, promote chemical reactions with the forming nanostructures, altering stoichiometry and creating NPs surrounded by carbon shells[41–44]. The formation of amorphous CrMnFeCoNi and CrMnFeCoNiMo NPs, in addition to minor fractions of the fcc phase, through LSPC in acetonitrile has been previously reported[36]. A significant carbon doping in the NP volume and the confinement effect created by the carbon shell were hypothesized to stabilize the amorphous structure. However, a mechanistic understanding of the particle formation mechanisms and solvent-metal interactions influencing the characteristics of the generated amorphous HEA NPs is missing.

Therefore, this study deploys a multi-microscopy and spectroscopy approach to report on the structure and morphology of the carbon-containing $CrMnFeCoNiC_x$ HEA nanoparticles, along with their size-dependent composition distribution and thermal stability. We investigate the synthesis and *in situ* carbon doping of HEA NPs by reactive nanosecond-pulsed laser ablation in acetonitrile, ethanol, and acetone. This allows us to report on the systematic influence of the solvent-metal interactions during laser synthesis on the resulting characteristics of the HEA NPs using transmission electron microscopy (TEM) combined with selected area electron diffraction (SAED), scanning TEM (STEM) paired with energy dispersive X-ray spectroscopy (EDS), atom probe tomography (APT) and X-ray photoelectron spectroscopy (XPS). The produced particles are found to be amorphous with different degrees of carbon shell thickness depending on the solvent used. In-situ TEM heating studies demonstrate that carbon supersaturation is the rationale behind the amorphous phase formation, as it kinetically hinders the crystallization during particle formation. A mechanistic view of the particle formation during reactive laser ablation is proposed, and the impact of the relative kinetics of competing reactions on the formation of carbon shells and the compositional distribution is discussed. Our results evidence a solvent-driven structural and compositional control of amorphous, carbon-doped HEA nanoparticles, with LSPC proving its potential for the synthesis of compositionally complex nanostructures in high volume.

## Results and Discussion

### Characteristics of nanoparticles synthesized in acetonitrile

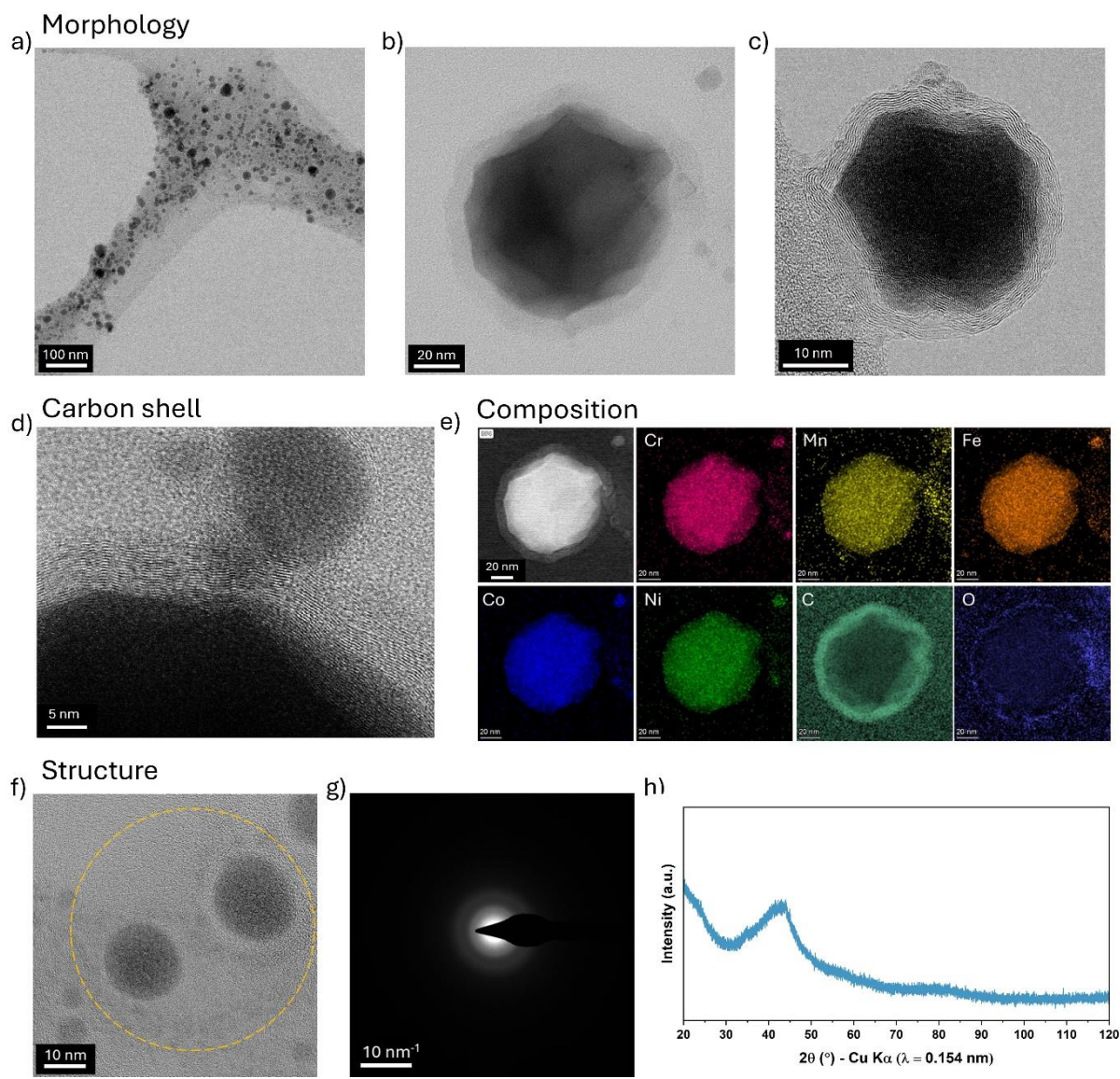

*Figure 1 Morphology, composition, and structural characterization of HEA NPs synthesized in acetonitrile. a) TEM bright field image showing the nanoparticles' morphology and size variations; b), c) STEM bright field images of selected nanoparticles of different sizes highlighting their rugged surface covered with a carbon shell around; d) STEM bright field image at a higher magnification showing the onion-like morphology and size-dependent carbon shell thickness differences between two NPs; e) STEM-EDS analysis showing the individual element distribution maps; f), g) TEM bright field image and its respective SAED pattern highlighting the amorphous nature of the NPs; h) Powder XRD pattern of the NPs showing a diffused peak indicating their amorphous nature.*

The colloidal HEA NPs synthesized by LSPC in acetonitrile were imaged by TEM, **Figure 1a**. Details of the structural and compositional characterization of the target material are discussed in the Supporting Information. The particle size ranges from 5 nm to hundreds of nanometers with an average diameter of 17 ± 5 nm (**Figure S1**). The HEA NPs synthesized in acetonitrile exhibit an irregularly shaped, rugged morphology (**Figures 1b–c**), which contrasts with the spherical particles obtained through conventional laser ablation synthesis, suggesting distinct particle formation and growth mechanisms. Each particle is encapsulated by a graphitic, onion-like carbon shell, **Figure 1d**. Mapping by STEM-EDS reveals a near-uniform distribution of constituent metallic elements, **Figure 1e**. Carbon is predominantly in the outer shell but is found throughout the particle volume (**Figure S2**). The diffuse rings in the SAED pattern, **Figures 1f and g**, indicate amorphous NPs, further confirmed by XRD of the dried colloid, **Figure 1h,** with a single broad peak at $2\theta \approx 43°$.

The thermal stability and microstructural evolution of these NPs were evaluated by *in-situ* TEM (see Methods), and **Figures 2a–b** show variations observed before and after heating up to 600˚C. Crystallization started in smaller particles (< 10 nm in diameter) during the heating step of 350 to 400 °C (**Figure S3**). At higher temperatures, crystallites nucleate in the larger particles and then grow till 600 °C, evidenced by the enhanced contrast, **Figures 2 b-i and iii**. Rings in the SAED before and after heating, **Figures 2a-iii and b-iii**, evidence polycrystalline, face-centered cubic particles, along with a Mn-rich oxide phase. During heating, the onion-like graphitic carbon shell thickness increased from ~10 to ~15 layers after reaching 600 °C, driven by the outward diffusion of carbon from the bulk of the NP towards the shell. The equilibrium solubility of carbon in the lattice decreases due to crystallization and segregates to the surface. From thermodynamic modeling, the equilibrium C solubility in the face-centered cubic phase of CrMnFeCoNi alloy reported is relatively low at ~0.1 at.% at 1000 °C[45]. Sharp edges of crystallites marked by yellow arrows in **Figure 2b-ii** indicate that the exclusion of carbon is essential to enable crystallization, at the expense of increasing the surface energy, highlighting the critical role of carbon supersaturation[41,46] in pulsed laser synthesis in organic solvents.

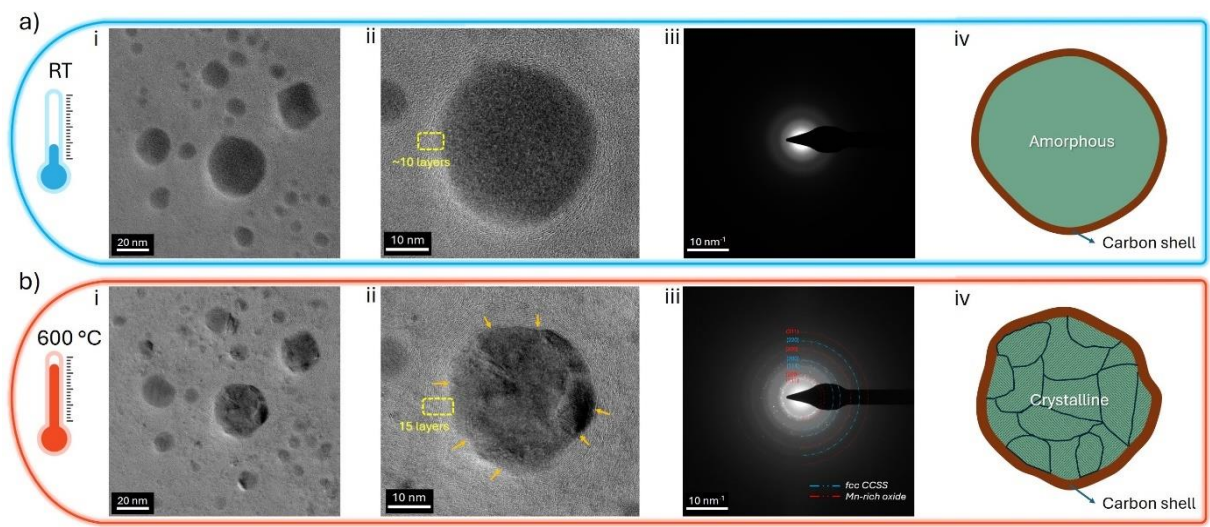

*Figure 2 In-situ TEM heating studies of HEA NPs synthesized in acetonitrile. a) TEM characterization of as-synthesized NPs: i) TEM bright field image of a selected NPs region showing their amorphous state up to 350°C (see Fig. S3), ii) TEM bright field image highlighting the carbon shell morphology around a selected NP, iii) SAED pattern of the NP shown in (ii) highlighting its amorphous nature, and iv) a schematic representation of the NP in its as-synthesized condition. b) TEM characterization of in-situ heated NPs at 600 °C: i) TEM bright field image of a selected NPs region showing their transition to a crystalline state, ii) TEM bright field image highlighting the evolution of carbon shell thickness around a selected NP, iii) SAED pattern of the NP shown in (ii) highlighting its crystalline nature after heating to 600 °C and iv) a schematic representation of the evolution of the NP after heating to 600 °C.*

**Size-dependent composition distributions of nanoparticles in acetonitrile**

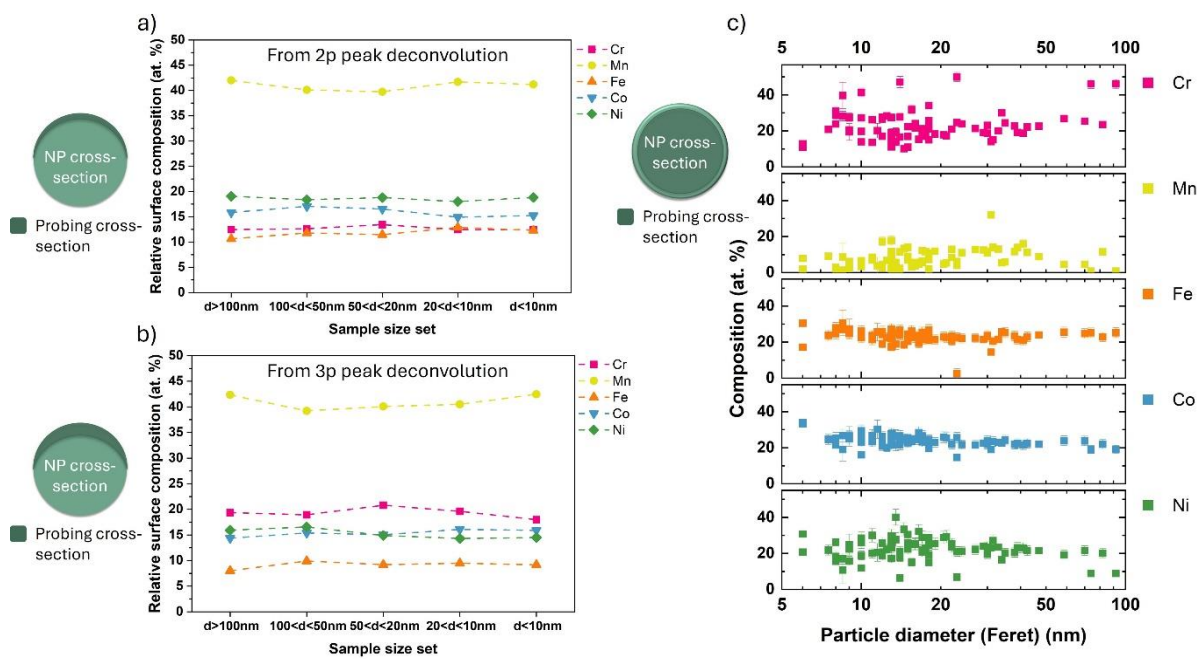

*Figure 3 Size-dependent composition distributions in HEA NPs synthesized in acetonitrile. a), b) Relative XPS surface elemental compositions as a function of different particle diameter*

*sub-sets calculated from the high-resolution XPS 2p and 3p spectra of the respective elements. c) Average elemental compositions as a function of particle diameter obtained from STEM-EDS. The schematic representation on the left of the plots shows relative probing volume cross-sections compared to the NP cross-sections of the respective measurement variant.*

By centrifugation, the as-synthesized amorphous HEA NPs of different sizes were separated into five classes (see methods). Quantitative analysis of XPS spectra gives the relative composition integrated over a few nanometers below the particle's surface ~4 nm and ~6–7 nm for 2p and 3p signals, respectively[47–49] (see methods). The surface composition is consistently offset from the equimolar composition of the target (**Figures 3a–b**). The concentration of Mn exceeds 40 at.%, whereas that of Fe is only 7–10 at.%, and Co and Ni 13–17 at.%. The concentration of Cr is around 13 at.% from deconvoluting the 2p signals and 20 at.% for 3p, indicating Cr segregation beneath a Mn-enriched surface. These measurements agree with previous computational studies on bulk crystalline Cantor alloys[50–52]. Transition metals, particularly Mn and Cr, are prone to oxidation in the conditions of laser synthesis[38], which involve supercritical vapor and liquid phases of the metallic constituents along with active species like carbon, oxygen, and hydrogen in the vicinity of the ablation plume[26]. Oxidation is expected to be minimized in organic solvents compared to aqueous media, yet it can drive a deviation from stoichiometry in the condensed NPs. STEM-EDS mapping (**Figures 1e and S2**) of individual NPs shows little to no detectable oxygen within the particle volume. Hence, the Mn-enriched surface cannot be attributed to oxidation during ablation. A shift of the peaks in the XPS spectra from the metallic species indicates the presence of oxide/hydroxides from possible oxidation post-ablation.

**Figure 3c** plots the composition from STEM-EDS, averaged through the thickness of the particle, as a function of particle diameter (Feret). Mn depletion, typically above 10 at.%, is evidenced regardless of size. Control measurements were made to eliminate deviations from specimen preparation or plasma cleaning of the grids prior to analysis (see Methods). Fe and Co concentrations remain near-stoichiometric. Cr and Ni concentrations show considerable scatter, particularly in NPs below 20 nm, while approaching stoichiometric values in larger particle sizes. The Cr enrichment observed in nanoparticles smaller than 20 nm can be attributed to its pronounced sub-surface enrichment tendency, whereas Ni exhibits the opposite trend, as also evidenced in the XPS composition profiles. STEM-EDS mappings, **Figures S4–S8**, show Mn-enriched fragments outside the NP regions and attached to the carbon shells, which contribute to surface Mn-enrichment in XPS. The Mn loss hence occurred during the ablation process, before the formation of the carbon shells, and the co-location of Mn and O in **Figure 1e** suggests the Mn-fragments formed oxides, likely during storage.

Complementary compositional analysis was obtained by APT, performed on a NP assembly and facilitated by in-situ Cr- coating[53], with the corresponding 3D reconstruction displayed in **Figure 4a**. The use of Cr for the coating prevents us from analyzing the distribution of Cr in the NPs themselves. **Figure 4b** is a close-up on the carbon shell surrounding the NPs highlighted by a brown isosurface, delineating regions containing more than 1.5 at.% C, surrounding NPs mapped by a blue isosurface delineating concentration greater than 20 at.% of Fe, Co, and Ni combined. The complex elemental environment with substantial C content (up to 20 at.%) in these NPs can induce local magnification and trajectory aberrations due to non-uniform evaporation fields, complicating the delineation of individual NPs[54–56]. A 1D composition profile calculated along a cylindrical region of interest (ROI - yellow), plotted in **Figure 4c**, provides the composition distribution between two NPs. 12 – 15 at.% of C is found within the NPs, and while Ni, Co, and Fe are relatively uniformly distributed, Mn is depleted. A carbon-enriched shell appears between the two particles, and the peak in Mn concentration is interpreted as Mn-rich fragments on the surface, as otherwise observed by STEM-EDS and XPS.

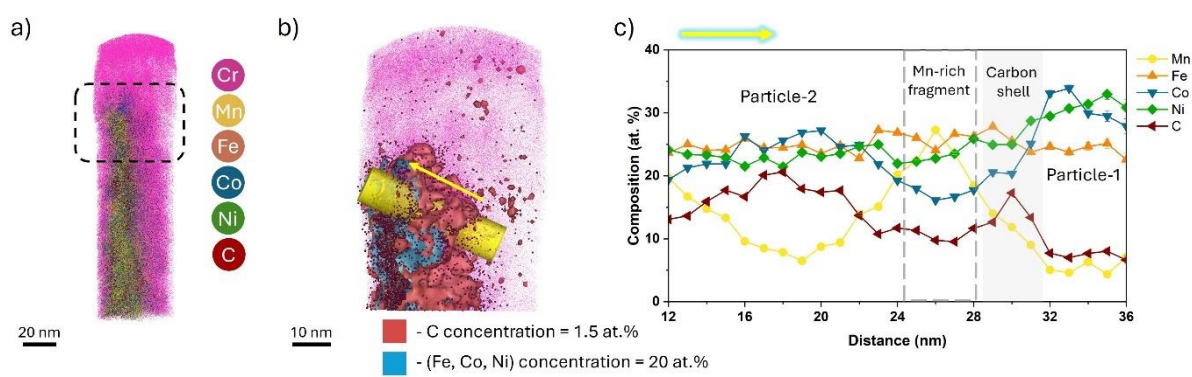

*Figure 4 APT analysis of HEA NPs synthesized in acetonitrile. a) 3D APT reconstruction of a NP assembly encapsulated in an in-situ coated Cr layer; b) Atom map from a selected region highlighting the carbon shell surrounding the NP region using an isosurface (brown) of C with a concentration greater than 1.5 at.%; c) 1D composition profile from the selected ROI shown in (b) highlighting the composition distribution as a function of distance.*

**Insights into amorphous nanoparticle formation**

Collectively, the compositional analyses by EDS, XPS, and APT agree and provide a thorough basis to rationalize the formation of amorphous HEA nanoparticles doped with carbon and with graphitic carbon shells surrounded by Mn-rich fragments in acetonitrile. **Figure 5** sketches our proposed mechanism. The laser pulse focused on the alloy target creates a plume of the target´s elements[26]. In multi-elemental targets, the distribution of each element within the plume is dictated by its physical properties, even when the target is a solid solution. For ultrashort-pulsed laser ablation in liquid, atomistic simulations predicted that rapid deceleration of the ablation plume by the liquid environment results in the formation of a transient hot and dense metal region at the front of the plume. This hampers the mixing of elements and contributes to the stratification of the plume in the emerging cavitation bubble[57]. The pressure at the shockwave front during nanosecond-pulsed laser ablation in organic solvents might also affect the intensity of plume-solvent vapor mixing[58]. Experimental identification of the main elemental mixing processes during laser ablation proved the benefits of longer laser pulses for mixing Fe and Co[59]. This motivated our use of nanosecond pulses herein, and it provided good and size-independent elemental mixing, except for Mn.

The slight loss of Mn during NP consolidation may be linked to thermophysical phenomena that are more prominent at longer pulse durations, which supply energy to the plume up to six orders of magnitude longer than in femto- or picosecond laser ablation[60]. Out of the five elements, Mn has the lowest latent heat of vaporization, the lowest melting point, and the highest vapor pressure, potentially leading to preferential separation within the ablation plume.

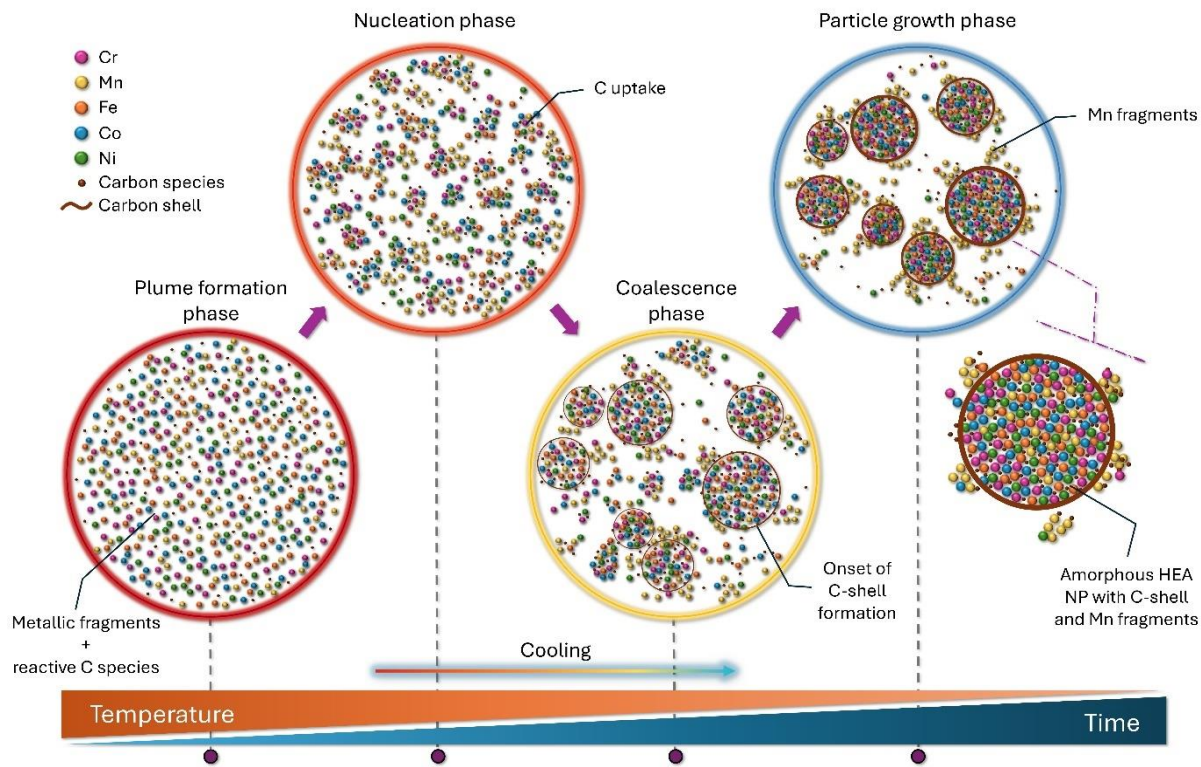

*Figure 5 Schematic illustration of the NP formation steps during laser ablation of CrMnFeCoNi target in acetonitrile leading to amorphous, carbon-doped HEA nanoparticles with graphitic carbon shells surrounded by Mn-rich fragments.*

Concurrently, plume-solvent interactions trigger the decomposition of organic molecules, generating active carbon species along with gases and byproducts[41]. This process creates an opportunity for interaction between ablated metallic fragments within the plume and the reactive carbon species, facilitating carbon uptake (**Figure 5**). The heat of the plume is transferred to the surrounding liquid, creating a vapor layer that encapsulates the ablated metallic content, initiating the formation of a cavitation bubble. Alloy NPs in the plume experience initial thermal quench rates of $10^{13}$ K/s, remaining above $10^{12}$ K/s until undercooling occurs at 1–2 ns, as observed in FeNi alloy ablation in water[61]. Larger NPs with slower cooling rates may require tens of ns to reach homogeneous crystallization thresholds, ultimately resulting in rapid solidification triggered by deep undercooling[61].

The presence of reactive metals like Cr, Fe, and Mn in the ablation plume, each with relatively high carbon solubility at their melting temperatures, respectively ~33, ~12, and ~7 at.%[42,62], promotes the retention of carbon species within the multimetallic plume. As cooling progresses, nucleation of the metallic species happens through the attachment of neighboring atoms. Two competing processes then occur simultaneously during condensation: (i) the coalescence of metallic fragments onto the growing nuclei within the plume[61], and (ii) the outward movement of excess carbon dissolved in the NPs towards their surface, initiating the formation of the carbon shell that can hinder further growth[41,46]. The observed rugged morphology of the formed NPs with outer carbon shells indicates the simultaneous occurrence of these competing processes. Mn, with its heterogeneous distribution within the ablation plume, would be the last element to integrate on the forming nanoparticle surface. Under rapid cooling, the excess dissolved carbon migrates to the NP's surface, initiating the formation of the carbon shell.  As the carbon shell begins to form, we could imagine that it competes with Mn coalescence, causing Mn to adhere to the developing carbon layer. With the subsequent collapse of the cavitation bubble, an onion-like graphitic carbon layer grows, supplemented by available carbon in the surrounding liquid and Mn-rich fragments attached[27,41] (**Figure 5**).

The plasma plume created by laser irradiation on the target surface exhibits extreme conditions, with initial temperatures in the order of $10^3$ K, pressures of $10^{10}$ Pa, and densities of $10^{22}$ atoms cm$^{-3}$ [63], cooling within a nanoseconds time scale[61]. Under such dynamic conditions, achieving thermodynamic equilibrium solidification becomes challenging, and the supersaturation of carbon into the metallic system could further impede the crystallization processes. Further studies supporting the formation of amorphous structures and enhanced glass-forming ability in alloys with the presence of carbon can be found[33,64,65]. A similar mechanism is observed in the case of HEA NPs in acetonitrile, where the rapid cooling conditions during ablation kinetically hinder the outward diffusion of carbon, thereby stabilizing an amorphous structure. Additionally, the incorporation of nitrogen, which is a known byproduct during ablation in acetonitrile[46], into the metallic constituents could further strengthen the formation of amorphous structures.

**Tuning carbon doping and carbon shell formation by organic solvent selection**

We extended our investigation to colloidal HEA NPs synthesized in acetone and ethanol to examine if these mechanisms can be extended to other organic solvents (see Supporting Information). Smaller NPs have rugged morphology and transition to regular spherical shapes in larger acetone-synthesized particles, while spherical particles are widely observed in ethanol (**Figures S9 and S12**). While HEA NPs synthesized in acetone maintain an amorphous structure, those in ethanol exhibit partial crystallization of the fcc phase (**Figures S9 and S12**). The carbon shell thickness decreases from acetonitrile to acetone to ethanol (**Figure S13**). XPS and STEM-EDS composition analyses reveal reduced Mn loss in particle volumes and near-stoichiometric distribution of other principal elements in both acetone and ethanol samples (**Figures S11 and S14**). From these observations, the particle formation mechanism rationalized for acetonitrile can be extended to acetone and ethanol in terms of carbon supersaturation and the respective dynamics of carbon shell formation and metallic coalescence during particle formation.

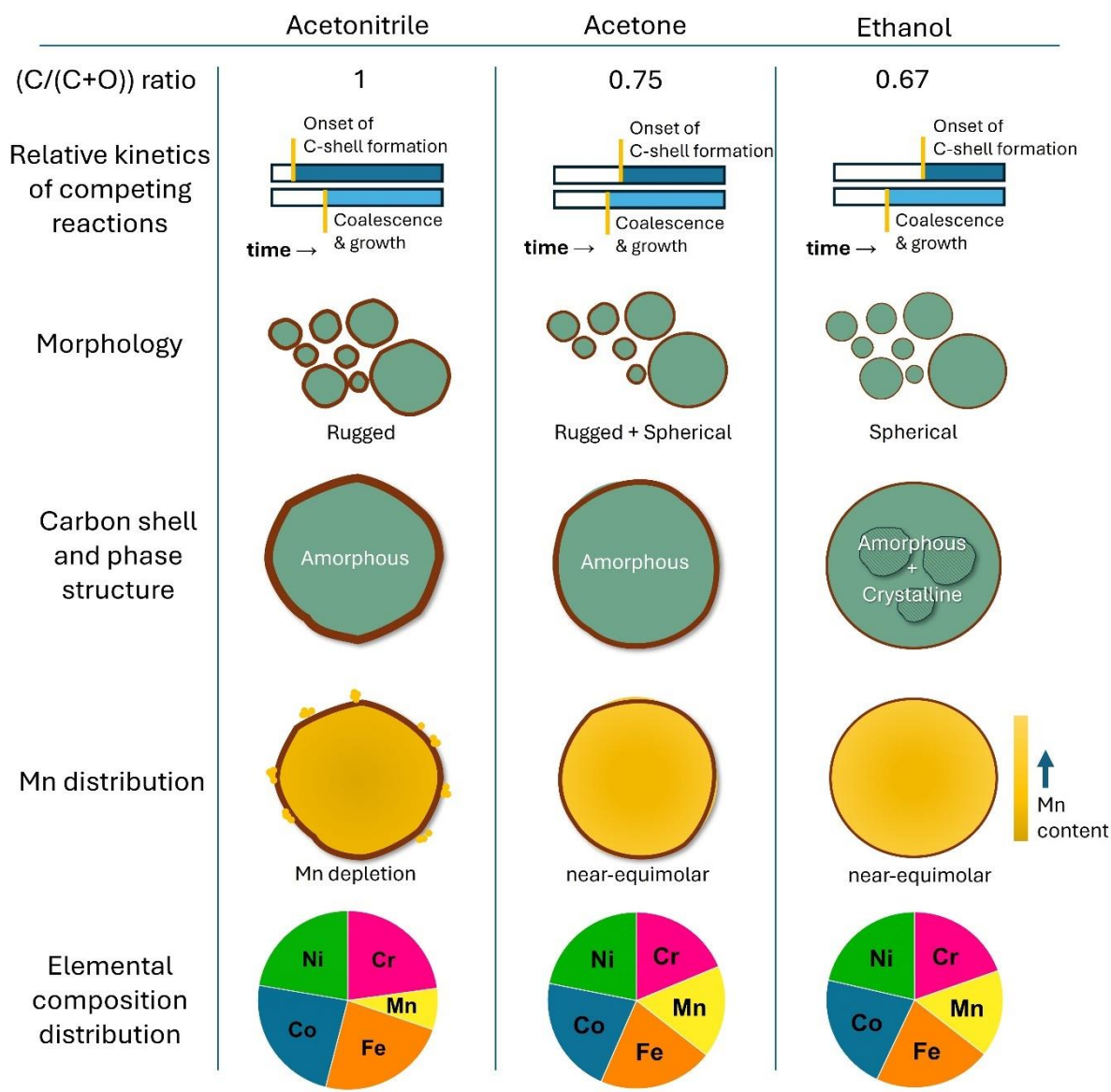

*Figure 6 Schematic illustration comparing the mechanistic changes, particle morphology, structure, and compositional distribution variations of HEA NPs synthesized in acetonitrile, acetone, and ethanol.*

**Figure 6** schematically compares the mechanistic changes, particle morphology, structure, and compositional distribution variations of HEA NPs synthesized in acetonitrile, acetone, and ethanol. A previous report on LPSC of Fe nanoparticles[66] had indicated an influence of the carbon fraction (C/(C+O)) in the solvent, which varies for ethanol (0.67), acetone (0.75), and acetonitrile (1.0) on the crystallinity of the formed NPs. Similarly, we observe less carbon incorporation during ablation in acetone and ethanol, indicated by thin carbon shell formation and insufficient carbon supersaturation to kinetically hinder crystallization compared to acetonitrile.

The carbon supersaturation in the ablation plume depends on the C/(C+O) ratio of the solvent and is hence a critical factor that mediates the competing processes of carbon shell formation and metallic constituent coalescence during nanoparticle condensation (**Figure 6**). In acetonitrile, where carbon supersaturation is highest, amorphous NPs with thicker carbon shells and rugged morphology are formed. The carbon shell formation dominates over the coalescence process, resulting in reduced Mn content, with the lost Mn found attached as fragments outside the shell (**Figure 5**). Acetone, with lesser carbon supersaturation, forms amorphous NPs with carbon shells of reduced thickness and produces a mixed morphology of larger spherical particles and smaller rugged ones. In this case, the metallic coalescence moderately precedes the carbon shell formation, preventing Mn loss. A similar effect is observed with ethanol, where partial crystallization occurs with thin carbon shells around the particles. Coalescence of metallic constituents dominates over carbon shell formation where Mn depletion is minimal, with other elements distributed close to the expected stoichiometric values. The ultrafast cooling of the ablation plume in the high fluence regime of pulsed laser synthesis quickly freezes the molten NP droplets and can explain why the composition is mostly independent of the particle size [61].

Solvent-dependent carbon doping via reactive laser ablation presents multiple tuning opportunities for synthesizing functional high-entropy nanoalloys. Despite HEAs often exhibiting size-dependent compositional variations and post-synthesis phase separations, pulsed-laser ablation enables the robust compositional design of nanoparticles independent of their size[67–69]. Amorphous phase formation as a result of carbon doping in such nanoalloys promotes their use in thermal catalysis up to temperatures of 350 °C with demonstrated enhanced catalytic activity and stability due to abundant unsaturated atomic coordination and fast electron transfer[33,70]. Additionally, hindered crystallization during particle formation prevents preferential phase separation or short-range ordering, thereby preserving the homogeneous chemical environment characteristic of multicomponent alloys. The carbon shell formation potentially improves conductivity during electrocatalysis while enhancing stability through a confinement effect that prevents elemental dissolution under reaction conditions[71]. Furthermore, the ability to modulate the Mn content on the particle's surfaces through solvent selection serves as a valuable tool in catalyst design, as surface Mn concentration has been identified as an activity descriptor for specific electrocatalytic reactions[21,72]. Finally, the kinetically driven particle formation occurring during laser ablation presents a notable advantage over other synthesis methods, as evidenced by the absence of preferential phase separation or island formation within the HEA NPs despite the substantial loss of one element.

**Conclusions**

We have demonstrated the synthesis of amorphous nanoparticles based on the Cantor alloy using pulsed laser ablation in selected organic solvents. The kinetic-control nature of the process provides high compositional robustness, independent of the particle size, as well as thermally quite stable particles. The interaction of reactive carbon species with the metallic plume at the nanoseconds-prolonged ultrahigh temperatures and the HEA´s supersaturated carbon content causes the stabilization of the amorphous structures, which is evidenced by in-situ TEM heating experiments. The solvent molecule determines the carbon doping, and part of the supersaturated excess carbon migrates to the surface during condensation to form the onion-like graphitic carbon shells. The degree of solvent-dependent carbon supersaturation alters the time scales of the competing reactions—carbon shell formation and metallic coalescence—thereby altering composition distribution and carbon shell thickness, which we mechanistically rationalized through a proposed particle formation mechanism. Additionally, reduced carbon availability for reaction with the ablation plume results in spherical morphology and partial crystallization in ethanol. Hence, during nanosecond-pulsed laser synthesis of these carbon-affine HEA colloids, solvent molecule selection allows for dictating the degree of amorphization (by carbon doping) and the carbon shell thickness. This provides a novel pathway for tuning nanoparticle morphology, structure, and surface composition via reactive LSPC.

## Methods

### Target preparation

The bulk equimolar CrMnFeCoNi HEA alloy was produced from pure metals by melting and casting using a vacuum induction furnace. Blocks of dimension $10 \times 20 \times 60$ mm$^3$ were machined from the cast block, which were then hot-rolled at 950 °C to reduce the thickness from 10 mm to 5 mm (50% reduction ratio). After the hot rolling process, the sheets were homogenized at 1200 °C for 3 hours in Argon atmosphere, followed by water-quenching. The sheets were then subjected to a cold rolling process to further reduce the thickness from 5 mm to 2.5 mm (50% reduction ratio). A final annealing step was carried out at 900 °C for 1 hour in Argon atmosphere, followed by water-quenching.

### Pulsed-laser ablation in liquids

The nanoparticle synthesis was performed using pulsed-laser ablation with a nanosecond EdgeWave GmbH laser with a wavelength of 1064 nm, pulse duration of 8 ns, repetition rate of 5 kHz, and pulse energy of 35 mJ. A polished bulk HEA target was positioned in a flow chamber, where distilled and nitrogen-bubbled organic solvents (acetonitrile, ethanol, or acetone) flowed continuously at approximately 50 mL min$^{-1}$. Colloidal nanoparticles were produced by scanning the focused laser beam across the bulk target in a rectangular pattern.

### Materials characterization

Qualitative phase analysis of the bulk target and nanoparticles was performed using X-ray diffraction (XRD) measurements (Rikaku Smartlab 9 kW diffractometer) with Cu Kα radiation ($\lambda = 0.15406$ nm), where the 2θ angle was varied between 20° and 120° with a step size of 0.01° and a scan speed of 1°/min. A voltage of 45 kV and a current of 200 mA were used during the measurements.

Composition analysis of the bulk target was performed using scanning electron microscopy (SEM) coupled with energy dispersive X-ray spectroscopy (EDS) (Zeiss Sigma). An accelerating voltage of 15 kV and a probe current of 7.7 nA was used.

The oxidation states and relative surface elemental composition analysis of the nanoparticles were performed by X-ray photoelectron spectroscopy (XPS) using a VersaProbe II™ from Ulvac-Phi using the Al-Kα line at 1486.6 eV and a spot size of 100 μm at an energetic resolution of 0.5 eV. A hemispherical analyzer (with an angle of 45° between the sample surface and the analyzer) and a dual-beam charge neutralization were used, whereby the detector was operated at a pass energy of 23 eV. By centrifugation, the as-synthesized amorphous HEA NPs of different sizes were separated into five classes (d>100 nm, 100<d<50 nm, 50<d<20 nm, 20<d<10 nm and d<10 nm), and drop cast onto a quartz substrate for XPS measurements. CasaXPS software was used to perform the peak deconvolutions with a Shirley[73] background applied to the individual peak fits. The acquired high-resolution XPS spectra were calibrated according to the binding energy of adventitious carbon (284.8 eV) determined during deconvolution of the C 1s spectrum of each sample[74]. The constraints used during the peak fitting of the 2p and 3p spectra of all elements and the respective peak fits of all samples can be found in the Supporting Information. Auger contributions from Fe, Co and Ni were considered during the 2p peak deconvolutions. Both $2p^{3/2}$ and $2p^{1/2}$ signals of Fe were fitted considering the Auger contribution from Co and Ni in its binding energy range[75]. The peak types were chosen generally as asymmetric Lorentzian for the elementary peaks and symmetric Gauss-Lorentzian for the oxide/hydroxide peaks. The peak areas calculated from the deconvolutions of the $2p^{3/2}$ signals were used for the surface composition quantification. Additionally, complementary composition quantification using the peak areas calculated from the 3p signals of the individual elements was performed. The kinetic energy and inelastic mean free path of the photoelectrons emitted by different orbitals differ[76,77], and the compositional information from 3p orbitals originates from relatively deeper below the surface compared to signals from 2p orbitals. Electrons from 2p orbitals have higher binding energies, resulting in lower kinetic energies after photoemission and consequently shorter inelastic mean free paths. Conversely, 3p electrons, with their lower binding energies, possess higher kinetic energies and longer inelastic mean free paths. Hence, the signals in the 3p region contain information from deeper below the surface compared to the 2p signals. Using Fe as an example, the depth probed by 2p electron signals is approximately 4 – 4.5 nm, while for 3p signals it is around 6 – 7 nm without accounting for elastic scattering[47–49]. Similar values can be assumed for other elements.

Structure and morphology analyses of the nanoparticles were performed using transmission electron microscopy (TEM) bright field (BF) imaging (Thermo Fischer Titan Themis G2) operated at 300 kV coupled with selected area electron diffraction (SAED). The TEM specimens were prepared by drop casting 5 µL of colloidal nanoparticles on a lacey-carbon grid followed by drying in air. The TEM grids were plasma-cleaned for 20 s before analysis. Composition analyses of the nanoparticles were performed using scanning transmission electron microscopy (STEM) combined with energy dispersive X-ray spectroscopy (EDS) (Thermo Fischer Titan Themis 300 – probe corrected). The microscope was operated in STEM mode at an accelerating voltage of 300 kV, camera length of 100 mm, beam convergence angle of 23.8 mrad, and for imaging, a bright field detector and a high-angle annular dark field (HAADF) detector were used. The Velox 3.6.0 software from Thermo Fischer Scientific was used for image and EDS acquisition. The Kα peaks of the elements Cr, Mn, Fe, Co, and Ni were used for elemental mapping and respective quantifications. A standard Cliff-Lorimer (K-factor) quantification, including absorption correction, was used for EDS quantification. Each individual EDS map was an accumulation of at least 300 frames of image size 1024 × 1024 pixels with dwell times varying between 10 to 20 µs to ensure sufficient intensity counts on the spectra for quantification.

**In-situ heating studies**

In-situ heating studies of the CrMnFeCoNiC$_x$ NPs synthesized in acetonitrile were performed using a Thermo Fischer Titan Themis G2 TEM instrument operated at 300 kV. An in-situ chip holder with amorphous SiN$_x$ windows (Nano-Chips Wildfire Double Tilt – DENSsolutions) was used for the analysis, where the colloidal NPs were drop-cast and air dried. The in-situ chip was plasma-cleaned for 20 s before analysis. The sample was heated to 600 °C in steps of 100 °C until 300 °C with a heating rate of 1 °C/s and in steps of 50 °C until 600 °C with a heating rate of 0.2 °C/s. At each temperature step, a holding time of 60 seconds was used to stabilize the sample before imaging was performed.

**Atom probe tomography**

A dual beam scanning-electron microscope - Ga-ion focused ion beam (SEM-FIB) microscope (Thermo Fischer Helios 600i) was used to prepare the needle-shaped atom probe tomography (APT) specimens from the bulk HEA target by in-situ lift-out method[78], followed by annular milling until the tip radius is less than 100 nm. The nanoparticle aggregates were embedded in a Ni matrix using co-electrodeposition on a Cu substrate to aid the preparation of APT specimens. Detailed specimen preparation steps can be found here[79]. The prepared needle-shaped APT specimens from the nanoparticle aggregate regions were in situ-coated with metallic Cr for enhanced yield[53,80]. APT measurements were carried out using a Cameca LEAP 5000 XR, with a base temperature of 60 K, pulsed UV-laser mode at an energy of 60 pJ per pulse, a target detection rate of 0.5 %, and a repetition rate of 125 kHz. Data reconstruction and post-processing were done using the software AP Suite by Cameca Instruments following the voltage-based reconstruction protocol.

## Acknowledgments

S.B. acknowledges funding by the Deutsche Forschungsgemeinschaft (DFG) - Project 277627168. V.N. is grateful for the financial support from the International Max Planck Research School for Interface Controlled Materials for Energy Conversion (IMPRS-SurMat), now International Max Planck Research School for Sustainable Metallurgy (IMPRS-SusMet), and the Center for Nanointegration Duisburg-Essen (CENIDE). The authors are grateful to Uwe Tezins, Christian Broß, and Andreas Sturm for their support at the APT, SEM, and FIB facilities and thank Philipp Watermeyer and Volker Kree for their support at the TEM facilities at the Max Planck Institute for Sustainable Materials. Further, the authors acknowledge Florian de Kock (University of Duisburg-Essen) and Benjamin Breitbach (Max Planck Institute for Sustainable Materials) for performing the XRD measurements. The authors thank the Interdisciplinary Center for Analytics on the Nanoscale (ICAN) at the University of Duisburg-Essen for access to the XPS facility. V.N. greatly appreciates Ezgi Hatipoğlu and Mathias Krämer at the Max Planck Institute for Sustainable Materials for their assistance and help with the FIB specimen preparation and APT analysis. V.N. acknowledges Theo Fromme, Jan Söder, and Robert Stuckert for the helpful discussions on laser synthesis in organic solvents and high entropy alloys.

## Declaration of competing interest

The authors declare that they have no known competing financial interests or personal relationships that could have appeared to influence the work reported in this paper.

## Data availability statement

The data that support the findings of this study are available from the corresponding authors upon reasonable request.

# Kinetically controlling surface atom arrangements in thermally robust, amorphous high-entropy alloy nanoparticles by solvent selection

Varatharaja Nallathambi[1,2], Se-Ho Kim[2,3], Baptiste Gault[2,4], Sven Reichenberger[1], Dierk Raabe[2*], Stephan Barcikowski[1*]

[1] Technical Chemistry I and Center for Nanointegration Duisburg-Essen (CENIDE), University of Duisburg-Essen, 45141 Essen, Germany

[2] Max Planck Institute for Sustainable Materials, Max-Planck-Str.1, 40237 Düsseldorf, Germany

[3] Department of Materials Science & Engineering, Korea University, Seoul, 02841, Republic of Korea

[4] Department of Materials, Royal School of Mines, Imperial College London, London SW72AZ, United Kingdom

* corresponding authors: stephan.barcikowski@uni-due.de, d.raabe@mpie.de

**Structural and compositional characterization**

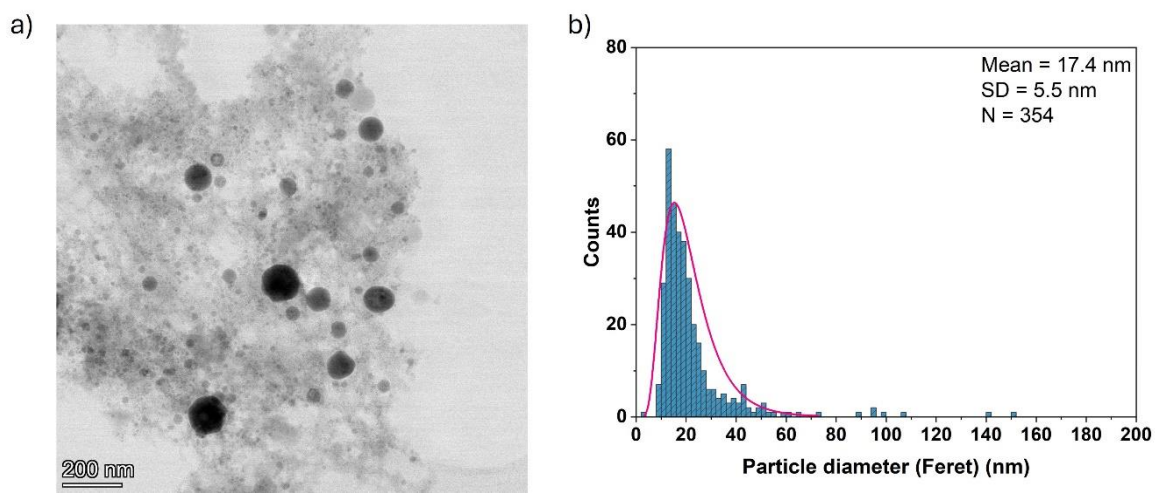

*Figure S7 a) STEM bright field image of the HEA NPs synthesized in acetonitrile and b) the corresponding size distribution (lognormal fit).*

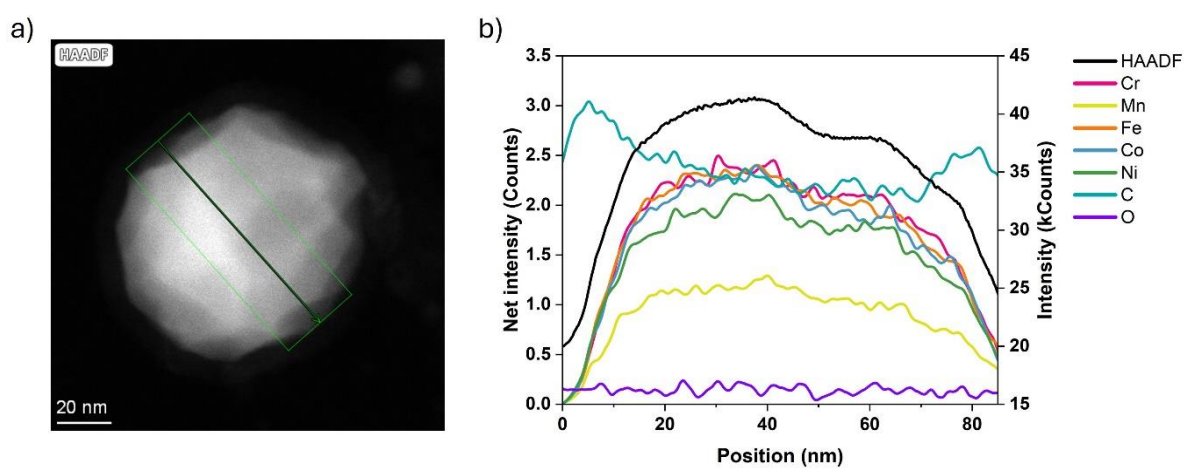

*Figure S8 a) STEM-HAADF image of a selected HEA NP synthesized in acetonitrile and b) the corresponding EDS line profile showing strong C signal intensity and very low intensity of O in the particle volume.*

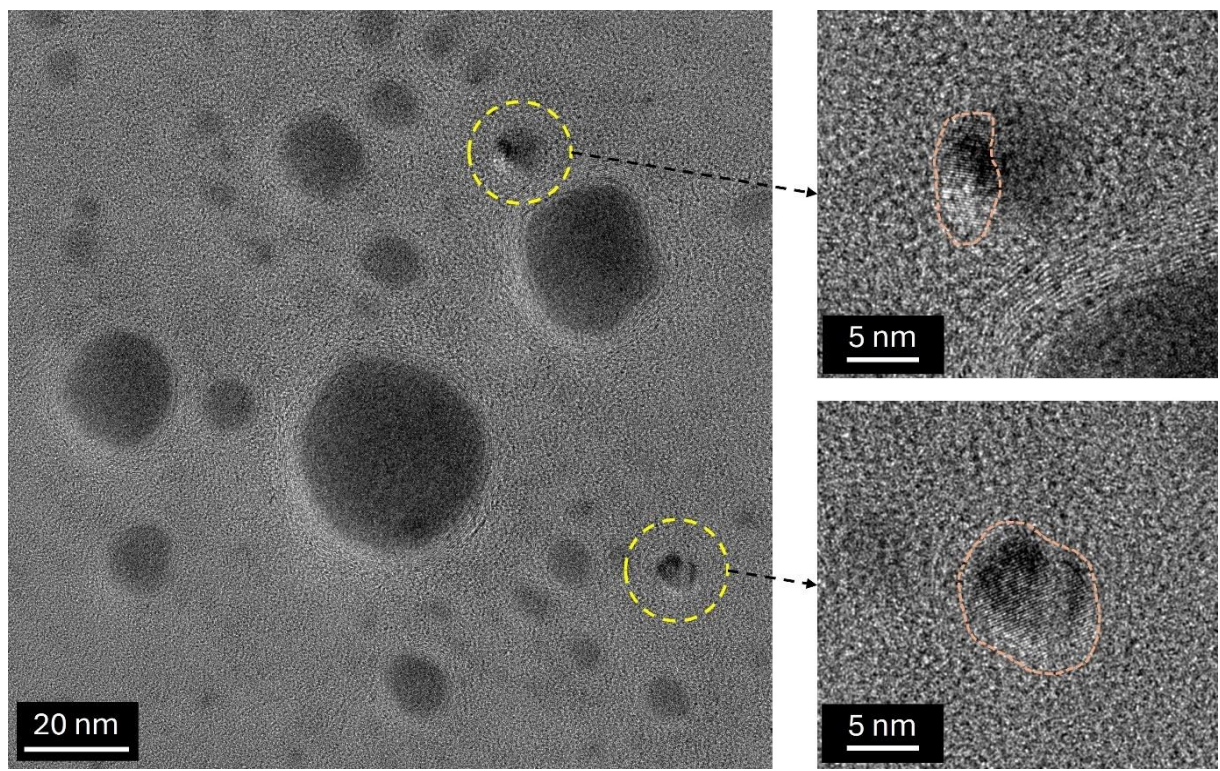

*Figure S9 TEM bright field images showing selected NP regions from the HEA NPs synthesized in acetonitrile, highlighting the start of crystallization during the in-situ heating step of 350 to 400 °C.*

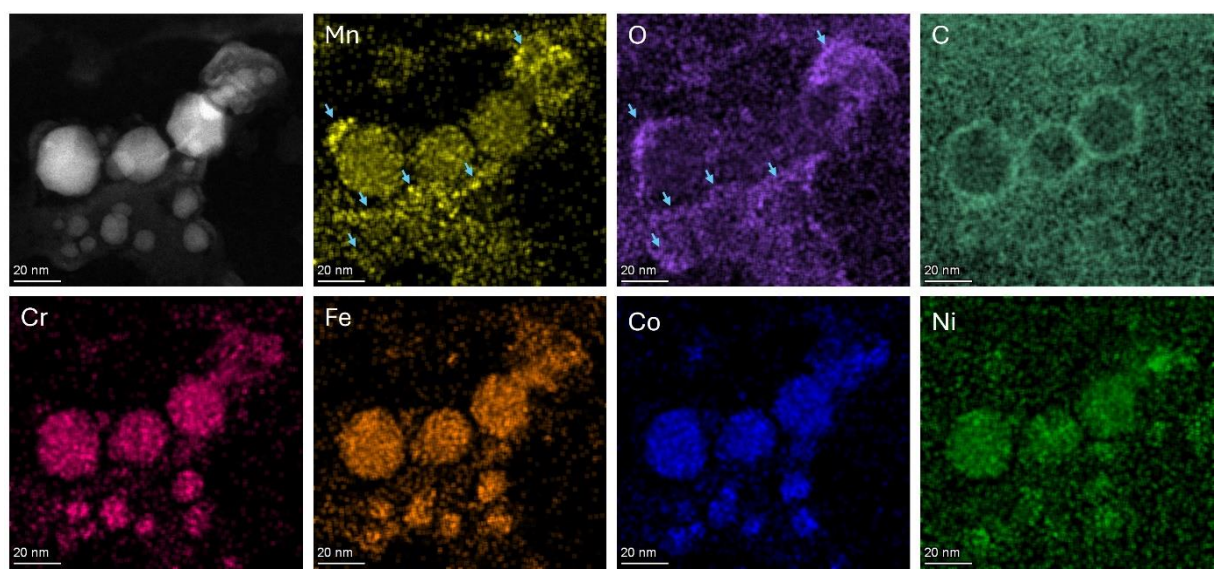

*Figure S10 STEM-EDS mapping of HEA NPs synthesized in acetonitrile, highlighting Mn-rich fragments (blue arrows) outside the carbon shell, which appear to be oxidized post-ablation.*

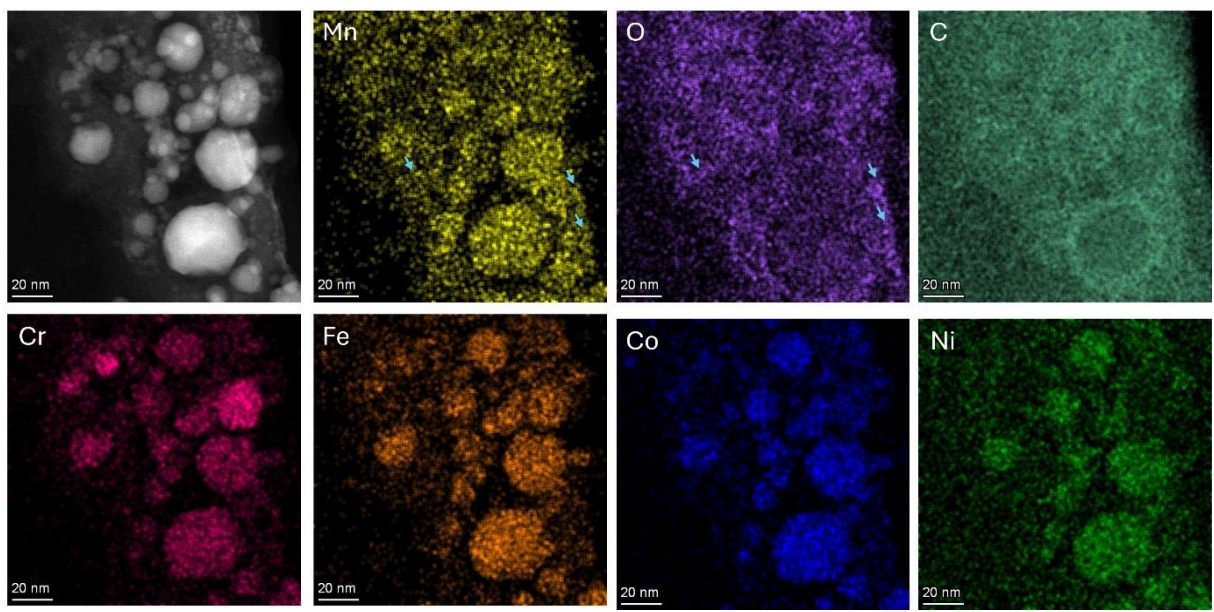

*Figure S11 STEM-EDS mappings of HEA NPs synthesized in acetonitrile, highlighting Mn-rich fragments (blue arrows) outside the carbon shell, which appeared to be oxidized post-ablation.*

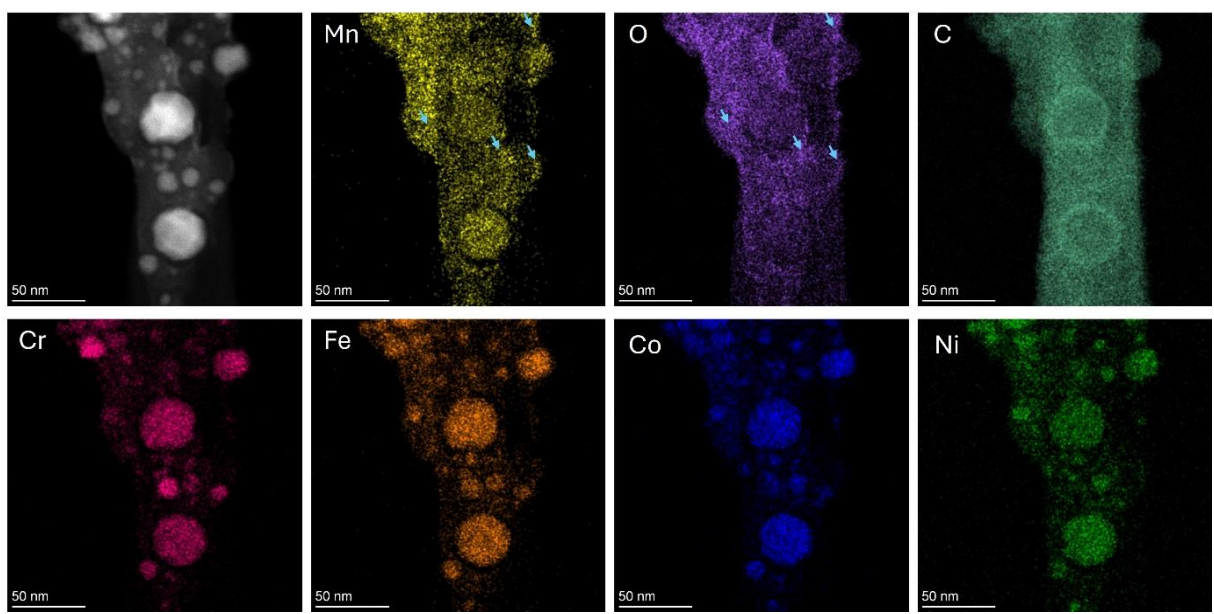

*Figure S12 STEM-EDS mappings of HEA NPs synthesized in acetonitrile highlighting Mn-rich fragments (blue arrows) outside the carbon shell which appeared to be oxidized post-ablation.*

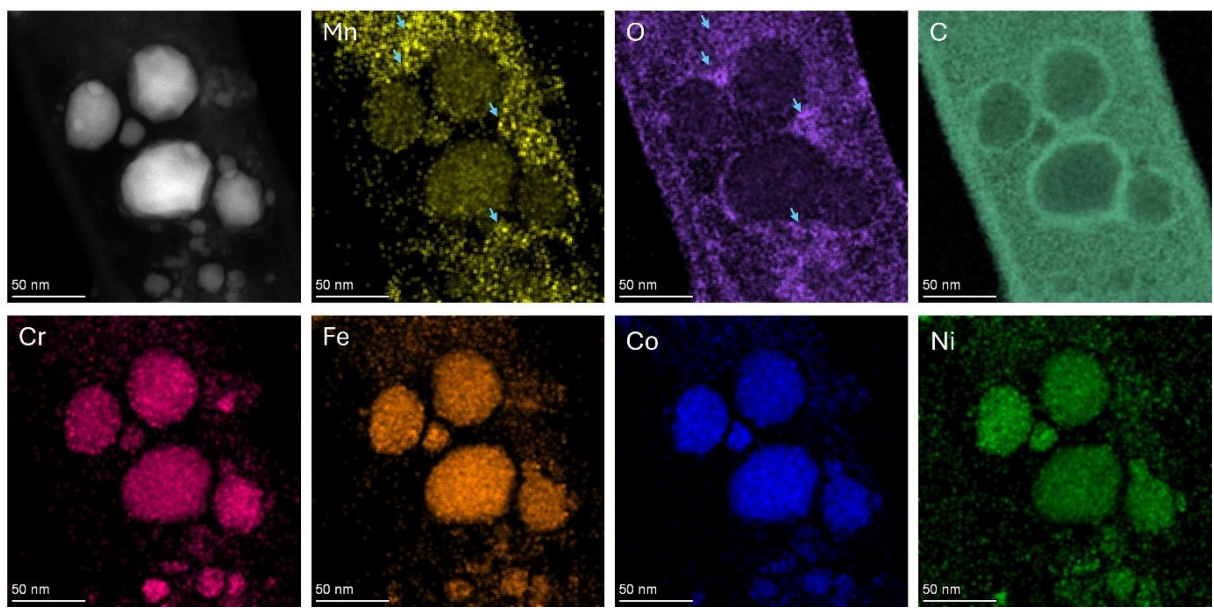

*Figure S13 STEM-EDS mappings of HEA NPs synthesized in acetonitrile highlighting Mn-rich fragments (blue arrows) outside the carbon shell which appeared to be oxidized post-ablation.*

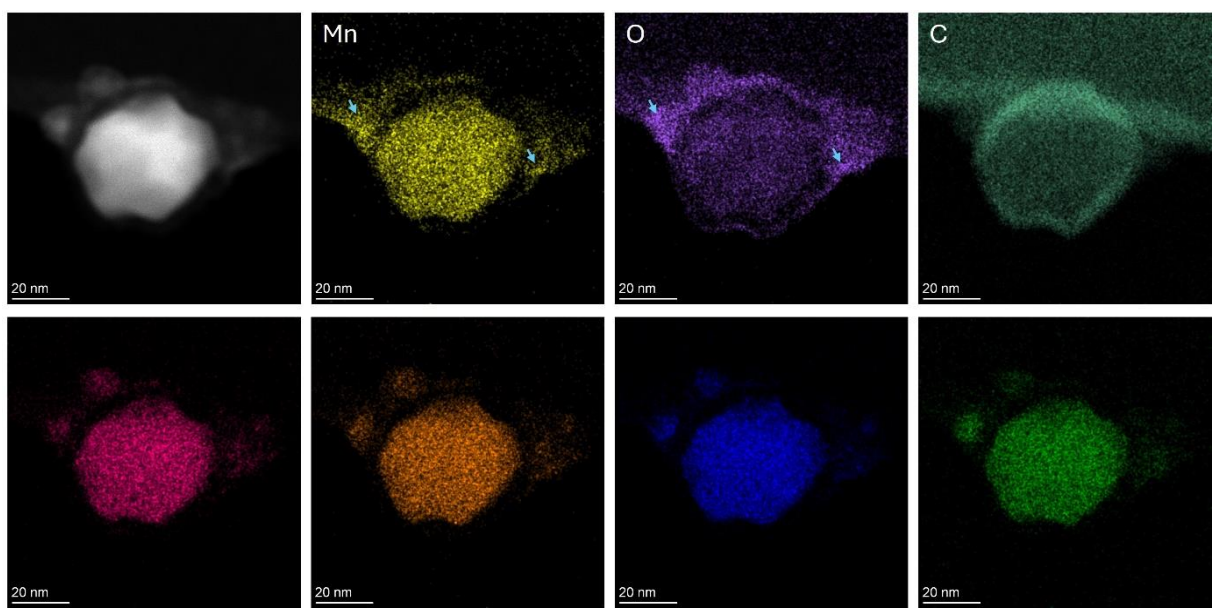

*Figure S14 STEM-EDS mappings of HEA NPs synthesized in acetonitrile highlighting Mn-rich fragments (blue arrows) outside the carbon shell which appeared to be oxidized post-ablation.*

**Characteristics of nanoparticles in acetone**

TEM characterization results performed for the NPs synthesized in acetone are shown in **Figure S9**. TEM and STEM bright field images (**Figures S9a – c**) reveal a nearly spherical morphology for larger particles, transitioning to rugged, for particles smaller than 40 nm. The particles display distinctive carbon shells, though these are less prominent compared to NPs synthesized in acetonitrile medium. This indicates a reduced carbon uptake during nanoparticle formation compared to the ablation process in acetonitrile as the excess carbon is expelled to the NP surface for the formation of onion-like graphitic carbon shells. STEM-EDS mappings (**Figure S9d**) show a uniform distribution of all five elements, with pronounced carbon signals in shell regions and minimal oxygen content. SAED analysis (**Figures S9e and f**) confirms the amorphous nature of the nanoparticles regardless of size, with additional supporting SAED data in **Figure S10**. Despite the comparatively lower carbon uptake than in acetonitrile, the carbon content within the nanoparticles remains sufficient to stabilize the amorphous phase.

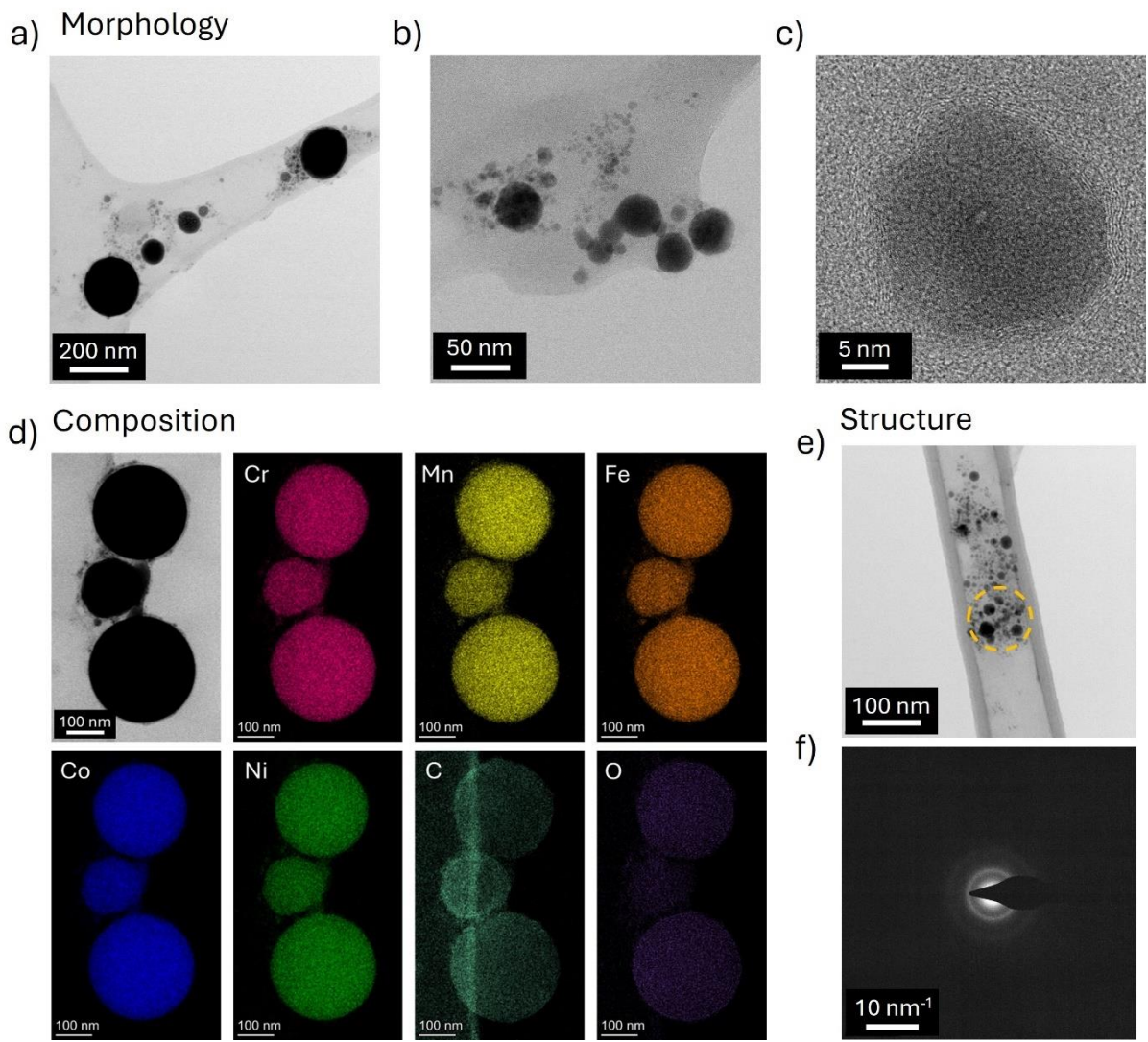

*Figure S15 Morphology, composition and structural characterization of HEA NPs synthesized in acetone. a) TEM; b), c) STEM bright field images showing the nanoparticles' morphology, size variations and carbon shell coverage; d) STEM-EDS analysis showing the individual element distribution maps; e), f) TEM bright field image and its respective SAED pattern highlighting the amorphous nature of the NPs.*

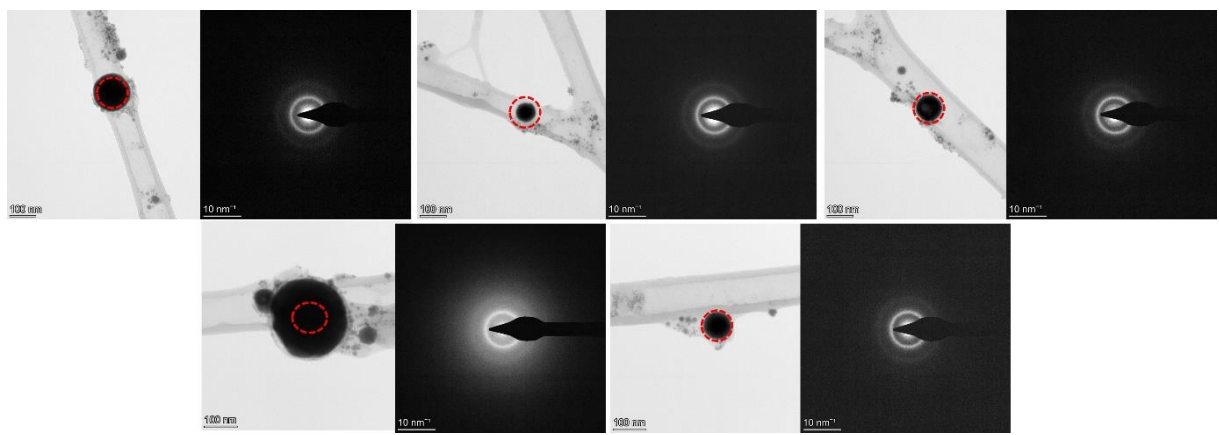

*Figure S16 TEM bright field images and their corresponding SAED patterns of HEA NPs synthesized in acetone.*

**Figures S11a and b** present XPS surface composition analysis for nanoparticles synthesized in the acetone medium with particle diameter, d greater than 100 nm. Attempts to calculate the surface composition from other size sets were unsuccessful due to poor signal-to-noise ratios of the respective 2p and 3p spectra. Similar to acetonitrile samples, Mn enrichment is observed in the surface layers, though at a reduced level (approximately 34 at.% compared to 42 at.% in acetonitrile samples). Cr also shows sub-surface enrichment, evidenced by higher content in calculations from 3p peak deconvolution versus 2p peak analysis. Co and Ni appear to be distributed close to stoichiometric values in surface layers, while Fe is slightly lower at 16 at.%. **Figure S11c** depicts composition values from individual particles as a function of particle diameter measured using STEM-EDS. Unlike acetonitrile samples, individual nanoparticle volumes show less severe Mn depletion, with an average value of 16 ± 2.5 at.% Mn across all size ranges. No significant particle-size-dependent compositional deviations were observed for any of the other elements, with only minor concentration scatter for Mn and Fe in a few fine particles below 30 nm. Otherwise, Cr, Fe, Co, and Ni maintain near-stoichiometric distribution throughout the analyzed size range with average values of 19 ± 1.8 at.%, 21 ± 2.1 at.%, 21.6 ± 1.3 at.% and 22 ± 2.2 at.%, respectively.

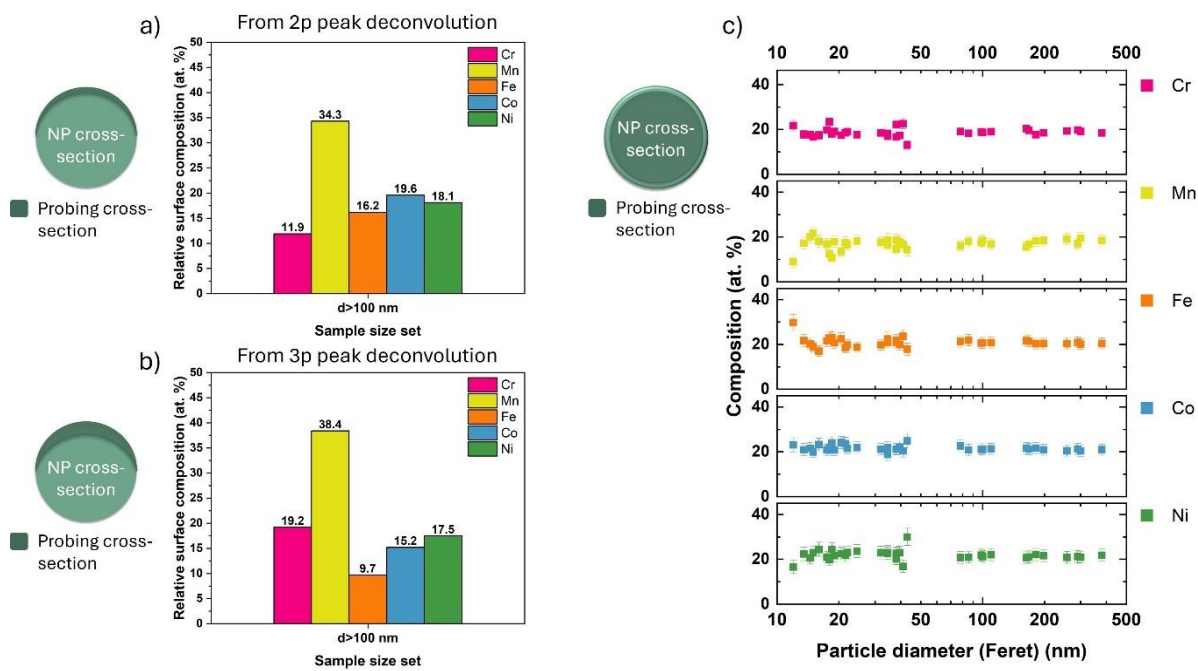

*Figure S17 Size-dependent composition distributions in HEA NPs synthesized in acetone. a), b) Exemplary relative XPS surface elemental compositions of sub-set with a particle diameter greater than 100 nm calculated from the high-resolution XPS 2p and 3p spectra of the respective elements of the NPs synthesized in Acetone (Note: The signal intensity of the XPS spectra of other particle diameter sub-sets were not sufficient for the calculation of relative surface composition, hence omitted); c) Average elemental compositions as a function of particle diameter obtained from STEM-EDS of the NPs synthesized in acetone. The schematic representation on the left of the plots shows relative probing volume cross-sections compared to the NP cross-sections of the respective measurement variant.*

**Characteristics of nanoparticles in ethanol**

**Figure S12** presents the morphology, composition, and structural characterization of HEA nanoparticles synthesized in ethanol. Most particles exhibit s spherical morphology, with rugged surfaces observed only in smaller particles below 15 nm diameter (**Figures S12a-c**). Particle surfaces display thin carbon shells that are less prominent compared to those in acetone and acetonitrile samples (**Figure S13**). STEM-EDS mappings reveal a uniform distribution of all elements with reduced carbon and oxygen signal intensity (**Figure S12d**). TEM-SAED analysis indicates primarily amorphous structures (**Figures S12e and f**), while powder-XRD analysis (**Figure S12g**) shows low-intensity (111) and (200) reflections of the fcc phase alongside the amorphous structure, suggesting partial crystallization in the NPs.

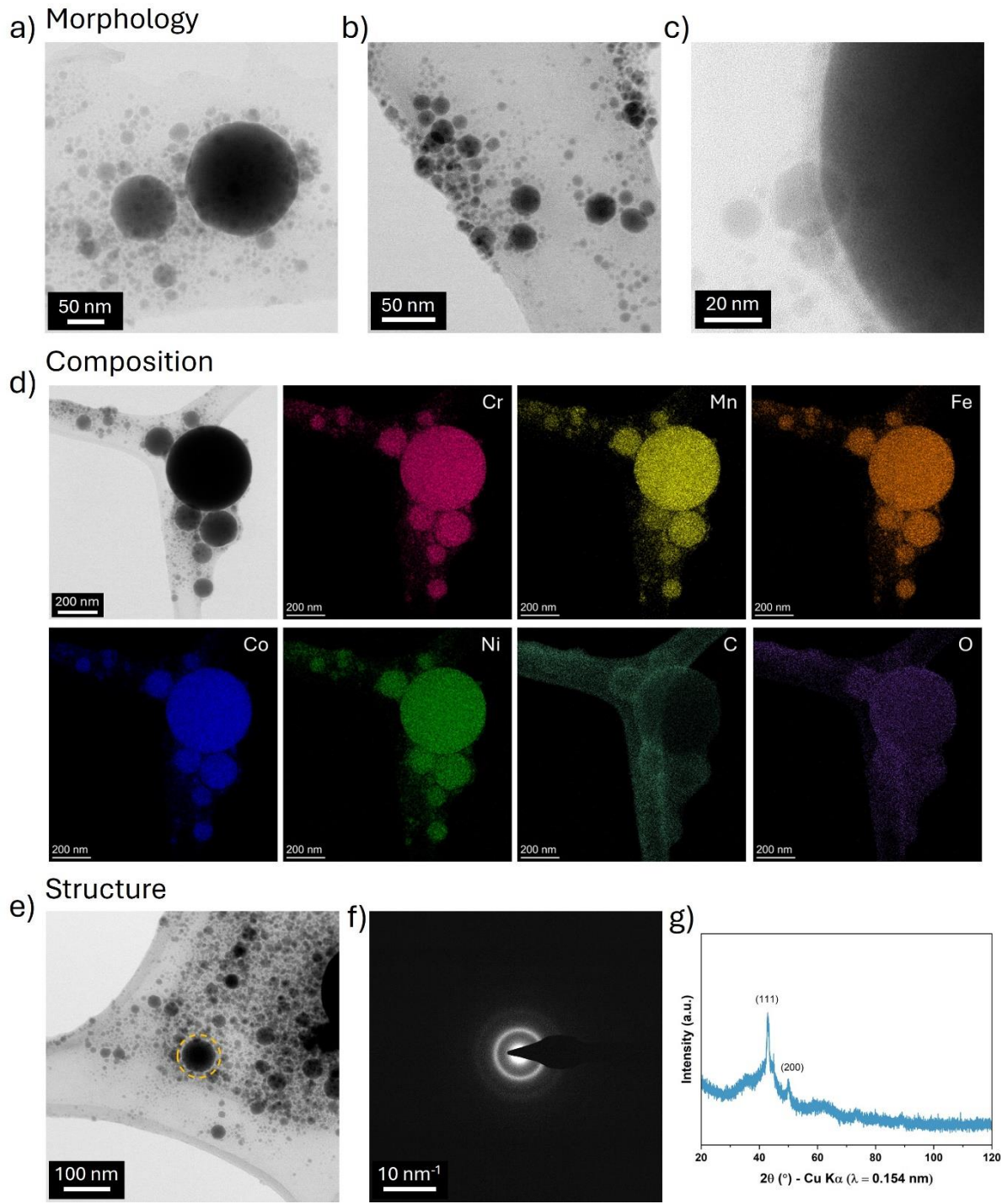

*Figure S18 Morphology, composition, and structural characterization of HEA NPs synthesized in ethanol. a), b), c) STEM bright field images showing the nanoparticles' morphology, size variations and carbon shell coverage; d) STEM-EDS analysis showing the individual element distribution maps; e), f) TEM bright field image and its respective SAED pattern highlighting the amorphous nature of the NPs; g) Powder XRD pattern of the NPs showing a broad diffused peak highlighting their amorphous nature with minor reflections indicating the presence of a small fraction of fcc solid solution phase.*

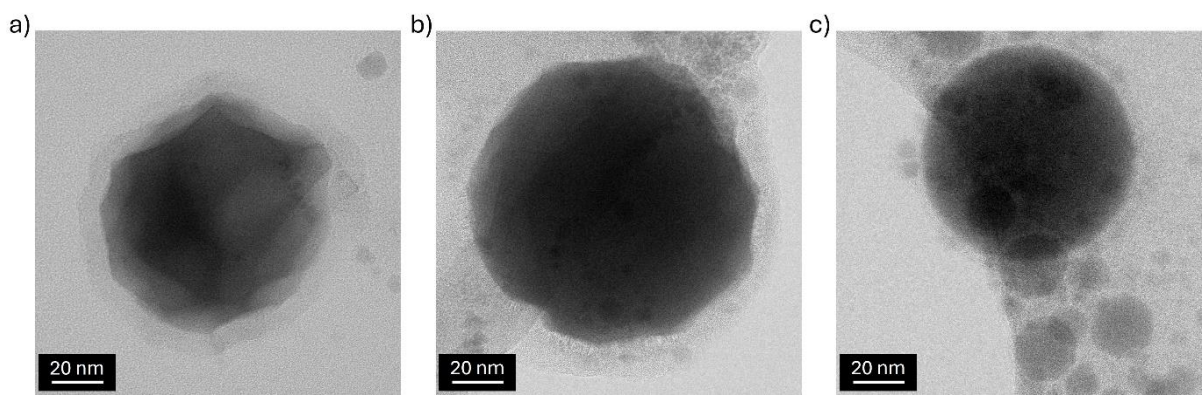

*Figure S19 STEM bright field images highlighting the differences in morphology and carbon shell thickness of HEA NPs synthesized in (a) acetonitrile, (b) acetone and (c) ethanol, respectively.*

The composition analysis of the HEA NPs synthesized in ethanol conducted using XPS and STEM-EDS is presented in **Figure S14**. Surface composition calculations derived from 2p and 3p peak deconvolutions across different size sets reveal Mn enrichment (**Figures S14a and b**), with a concentration of approximately 32 at.%, compared to 34 at.% in acetone samples and 42 at.% in acetonitrile samples. As observed in previous cases, sub-surface Cr enrichment is evident in the ethanol samples, while Fe, Ni, and Co content follows trends similar to those observed in acetone samples. **Figure S14c** illustrates average elemental composition values for individual particles as a function of particle diameter, determined via STEM-EDS. Fe, Co, and Ni exhibit uniform distributions throughout the analyzed particle size range, with average values slightly exceeding the expected stoichiometry: 21.3 ± 1.2 at.%, 21.6 ± 0.9 at.%, and 21.4 ± 1.2 at.%, respectively. Comparable to the results in acetonitrile, Mn loss within the particle volume is minimal, demonstrating near-uniform distribution across the size range. However, deviations in the concentration trend are observed for Mn in particles smaller than 30 nm in diameter, with corresponding minor adjustments in Cr and Fe concentrations. For particles exceeding 30 nm in diameter, the average Mn concentration is approximately 16 ± 1.9 at.%, while for particles below 30 nm, the value is around 17 ± 3.1 at.%. The average Cr concentration across all size ranges is 19.6 ± 1.6 at.%.

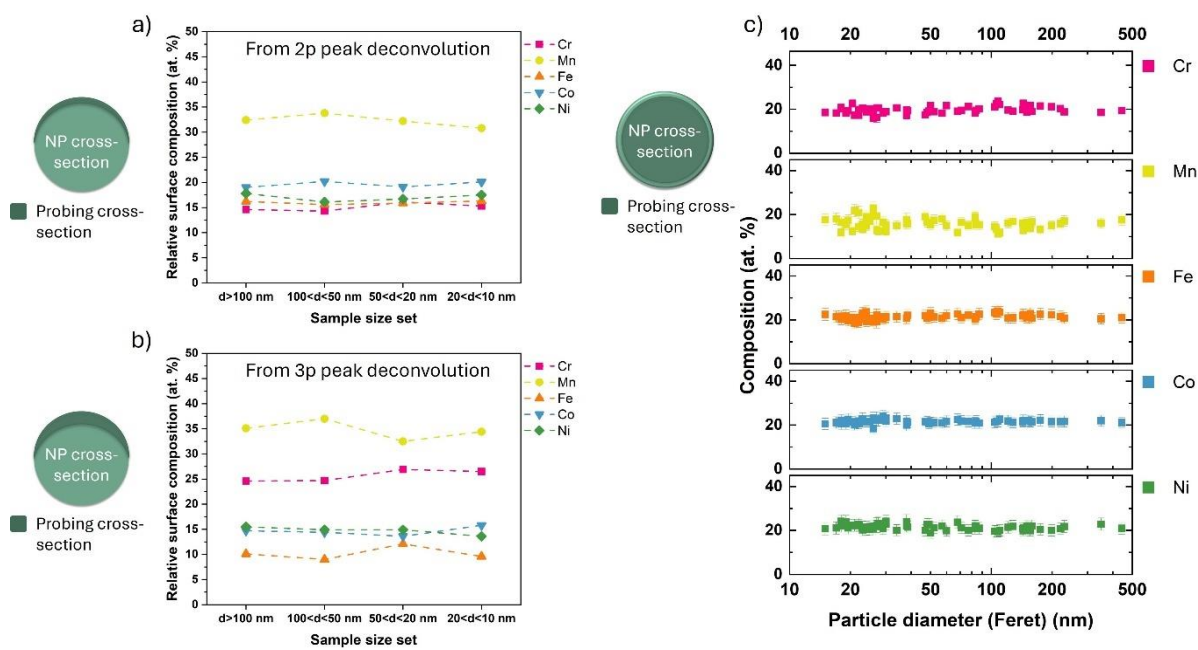

*Figure S20 Size-dependent composition distributions in HEA NPs synthesized in ethanol. a), b) Relative XPS surface elemental compositions as a function of different particle diameter sub-sets calculated from the high-resolution XPS 2p and 3p spectra of the respective elements of the NPs synthesized in Ethanol (Note: The signal intensity of the XPS spectra of the particle diameter sub-set less than 10 nm were not sufficient for the calculation of relative surface composition, hence omitted); c) Average elemental compositions as a function of particle diameter obtained from STEM-EDS of the NPs synthesized in ethanol. The schematic representation on the left of the plots shows relative probing volume cross-sections compared to the NP cross-sections of the respective measurement variant.*

**Target characterization**

XRD and SEM-EDS analysis were used to evaluate the phase structure and macroscopic elemental concentrations of the CrMnFeCoNi target material. **Figure S15a** shows the XRD pattern of the target indicating the peaks corresponding to a single-phase fcc crystal structure. **Figure S15b** shows the SEM-EDS mapping of a selected region on the polished target surface where no macroscopic element segregation was found. APT compositional analysis was performed to ensure the uniform elemental mixing at the nanoscale as preferential segregation in the target material would affect the resulting nanoparticle compositions. 3-dimensional (3D) atom maps of the constituent elements shown in **Figure S15c** show a uniform distribution, within the limits of resolution of APT [1,2]. 1D compositional analysis using a 45 nm-diameter cylindrical region of interest (ROI) from a selected region is shown in **Figure S15d** highlighting the near-equimolar and uniform distribution of the constituents. A spider plot with the average elemental compositions of the target material obtained from SEM-EDS and APT compositional analysis is shown in **Figure S15e** and the respective values are listed in **Table S1.**

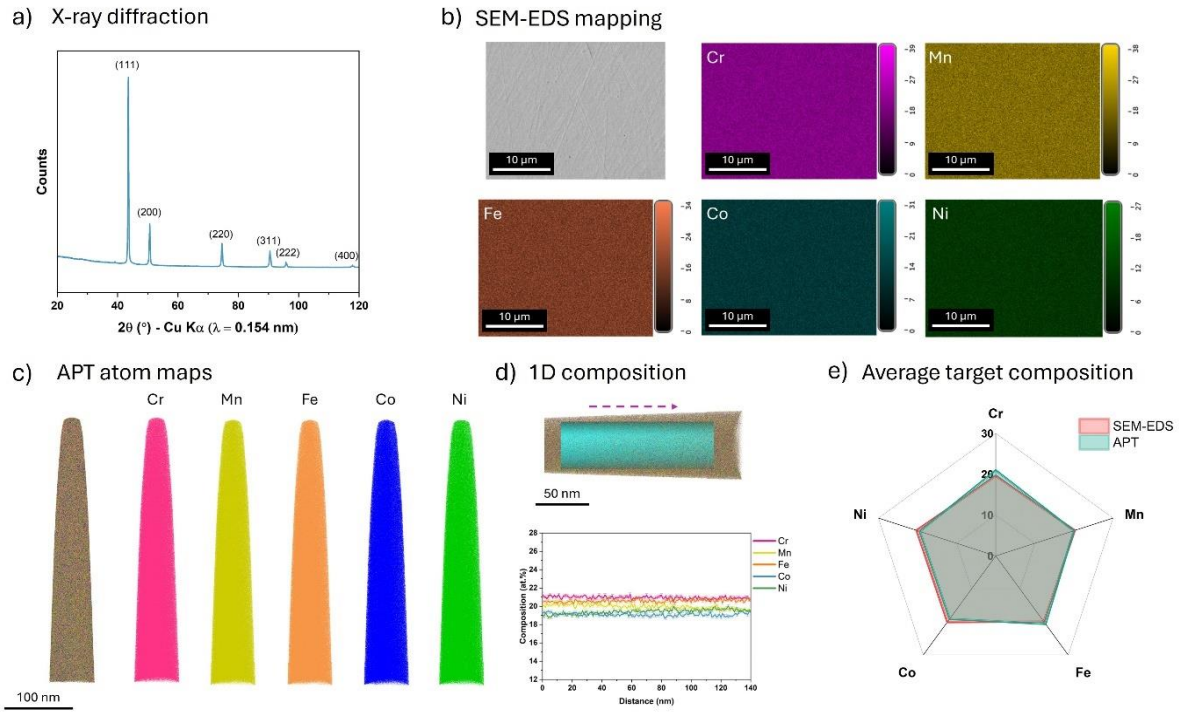

*Figure S21 Structural and compositional characterization of the CrMnFeCoNi bulk target. a) XRD pattern of the polished bulk target showing peaks corresponding to a single phase fcc crystal structure; b) SEM-EDS mapping of individual elements over a selected region on the polished bulk target surface; c) 3D APT reconstruction with individual element distribution maps of the bulk target showing no element segregations at the nanoscale; d) 1D concentration profile of individual elements from a selected region of the 3D reconstruction highlighting the near-equimolar stoichiometry of the target at the nanoscale; e) Spider plot showing the average elemental compositions of the target material obtained from SEM-EDS and APT compositional analysis.*

*Table S1 Average elemental concentrations from SEM-EDS and APT measurements of the CrMnFeCoNi bulk target.*

| Element | Average composition from SEM-EDS | | Average composition from APT | |
|---|---|---|---|---|
| | Atomic % | Error % | Atomic % | Error % |
| Cr | 19.59 | 2.9 | 20.95 | 0.15 |
| Mn | 20.22 | 2.9 | 19.95 | 0.14 |
| Fe | 19.91 | 3.1 | 20.59 | 0.15 |
| Co | 20.03 | 3.1 | 19.09 | 0.14 |
| Ni | 20.25 | 3.4 | 19.42 | 0.14 |

**XPS characterization**

*Table S2 Constraints used during the peak fitting of XPS 2p spectra listed for each element.*

| Cr 2p | $Cr^0$ | | Cr(III) oxide (5 multiplets) | | Cr(III) hydroxide | | Cr(VI) oxide | |
|---|---|---|---|---|---|---|---|---|
| | min | Max | min | max | min | max | min | max |
| Pos. Constr. / eV | 574 | 574.4 | 575.5 | 579.1 | 577.1 | 577.5 | 479.4 | 479.8 |
| FWHM Constr. / eV | 0.7 | 1.1 | 0.8 | 1 | 2.5 | 2.7 | 1.28 | 1.48 |

| Mn 2p | $Mn^0$ | | MnO (6 multiplets) | | MnOOH (5 multiplets) | | $MnO_2$ (5 multiplets) | |
|---|---|---|---|---|---|---|---|---|
| | min | Max | min | max | min | max | min | max |
| Pos. Constr. / eV | 638.5 | 639 | 640.2 | 645.9 | 641.1 | 645.5 | 642.4 | 646.8 |
| FWHM Constr. / eV | 1 | 1.3 | 1.5 | 1.95 | 1.05 | 1.45 | 1.05 | 1.45 |

| Fe 2p | $Fe^0$ | | FeO (5 multiplets) | | $Fe_2O_3$ (6 multiplets) | | FeOOH (6 multiplets) | |
|---|---|---|---|---|---|---|---|---|
| | min | max | min | max | min | max | min | max |
| Pos. Constr. / eV | 706.4 | 708 | 708.2 | 716.8 | 709.6 | 720.8 | 710.1 | 721.1 |
| FWHM Constr. / eV | 0.6 | 1 | 1.1 | 2.7 | 0.9 | 2.8 | 1.2 | 2.7 |

| Co 2p | $Co^0$ (3 multiplets) | CoO (4 multiplets) | $Co(OH)_2$ (4 multiplets) | CoOOH (4 multiplets) |
|---|---|---|---|---|

|  | min | max | min | max | min | max | min | max |
|---|---|---|---|---|---|---|---|---|
| Pos. Constr. / eV | 777.7 | 783.5 | 779.6 | 786.9 | 780 | 790.8 | 779.7 | 790.57 |
| FWHM Constr. / eV | 0.5; 2.8 | 0.9; 3.2 | 2.4 | 2.8 | 2.2 | 2.8 | 1.28 | 1.68 |

| Co 2p |  |  |  |  |  |  | Co$_3$O$_4$ (5 multiplets) |  |
|---|---|---|---|---|---|---|---|---|
|  |  |  |  |  |  |  | min | max |
| Pos. Constr. / eV |  |  |  |  |  |  | 779.2 | 789.9 |
| FWHM Constr. / eV |  |  |  |  |  |  | 1.1 | 1.7 |

| Ni 2p | Ni$^0$ (3 multiplets) |  | NiO (5 multiplets) |  | Ni(OH)$_2$ (6 multiplets) |  |
|---|---|---|---|---|---|---|
|  | min | max | min | max | min | max |
| Pos. Constr. / eV | 851.6 | 859.8 | 852.7 | 867.5 | 853.9 | 867.6 |
| FWHM Constr. / eV | 0.9; 2.28 | 1.1; 2.68 | 0.9; 3.1 | 1.1; 4.2 | 1.4; 2.85 | 1.7; 3.25 |

*Table S3 Constraints used during the peak fitting of XPS 3p spectra listed for each element.*

|  | Cr 3p |  | Mn 3p |  | Fe 3p |  | Co 3p |  |
|---|---|---|---|---|---|---|---|---|
|  | min | max | min | max | min | max | min | max |
| Pos. Constr. / eV | 46 | 47 | 47 | 50.4 | 53.7 | 56.7 | 59.9 | 62.5 |
| FWHM Constr. / eV | 0.6 | 5 | 0.6 | 5 | 0.6 | 5 | 0.6 | 5 |

|  | Ni 3p |  |
|---|---|---|
|  | min | max |
| Pos. Constr. / eV | 67.5 | 69 |
| FWHM Constr. / eV | 0.6 | 5 |

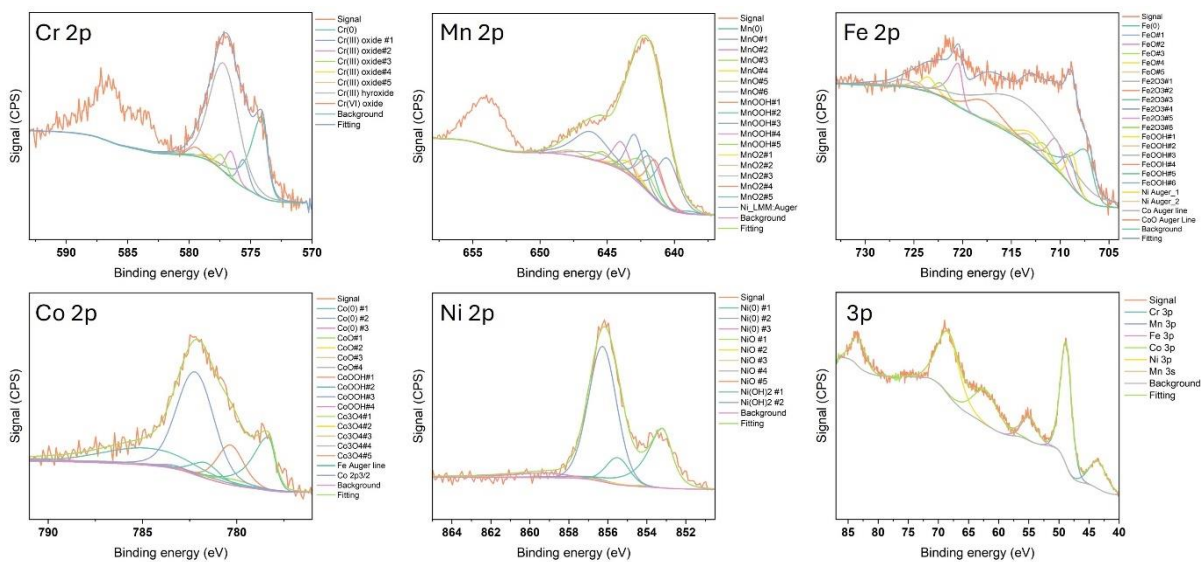

*Figure S22 Deconvoluted Cr 2p, Mn 2p, Fe 2p, Co 2p, Ni 2p and 3p XPS spectra of HEA NPs synthesized in acetonitrile within the size set d > 100 nm.*

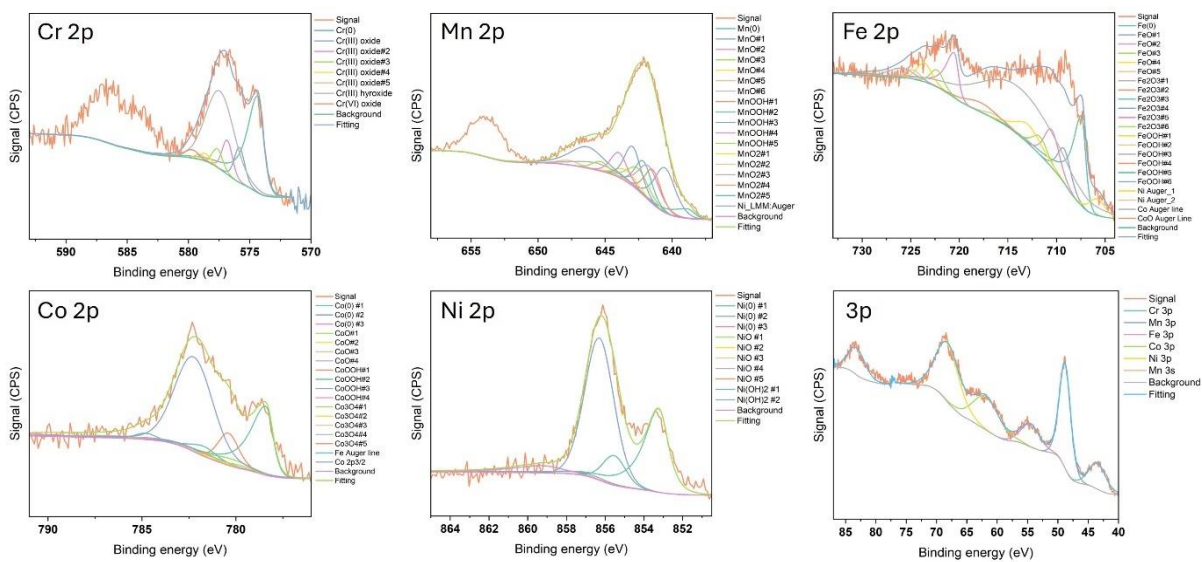

*Figure S23 Deconvoluted Cr 2p, Mn 2p, Fe 2p, Co 2p, Ni 2p and 3p XPS spectra of HEA NPs synthesized in acetonitrile within the size set 100nm < d < 50 nm.*

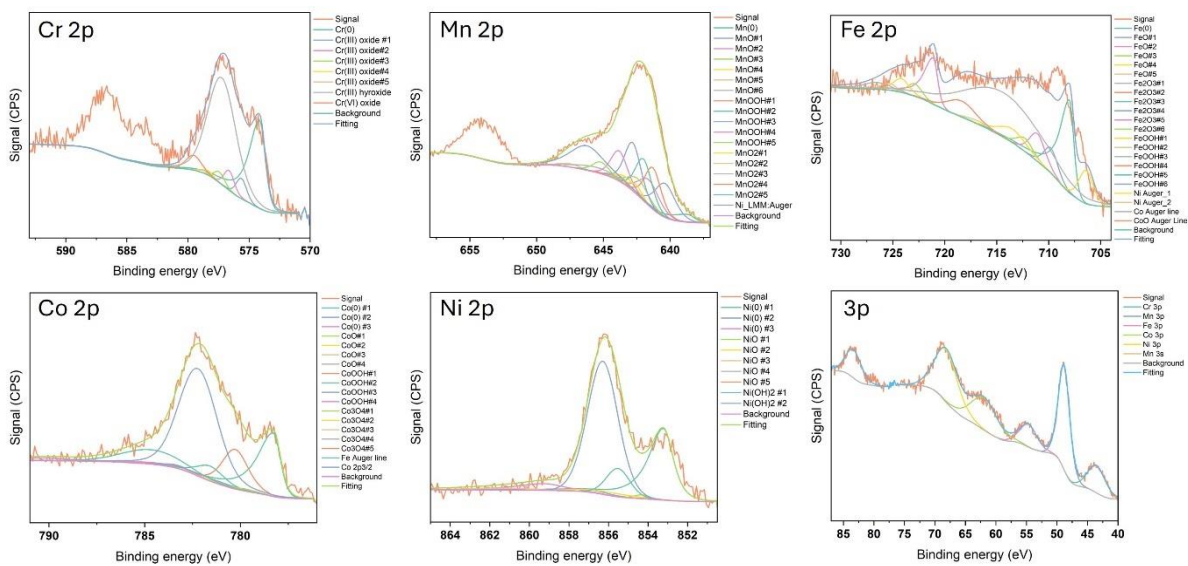

*Figure S24 Deconvoluted Cr 2p, Mn 2p, Fe 2p, Co 2p, Ni 2p and 3p XPS spectra of HEA NPs synthesized in acetonitrile within the size set 50nm < d < 20 nm.*

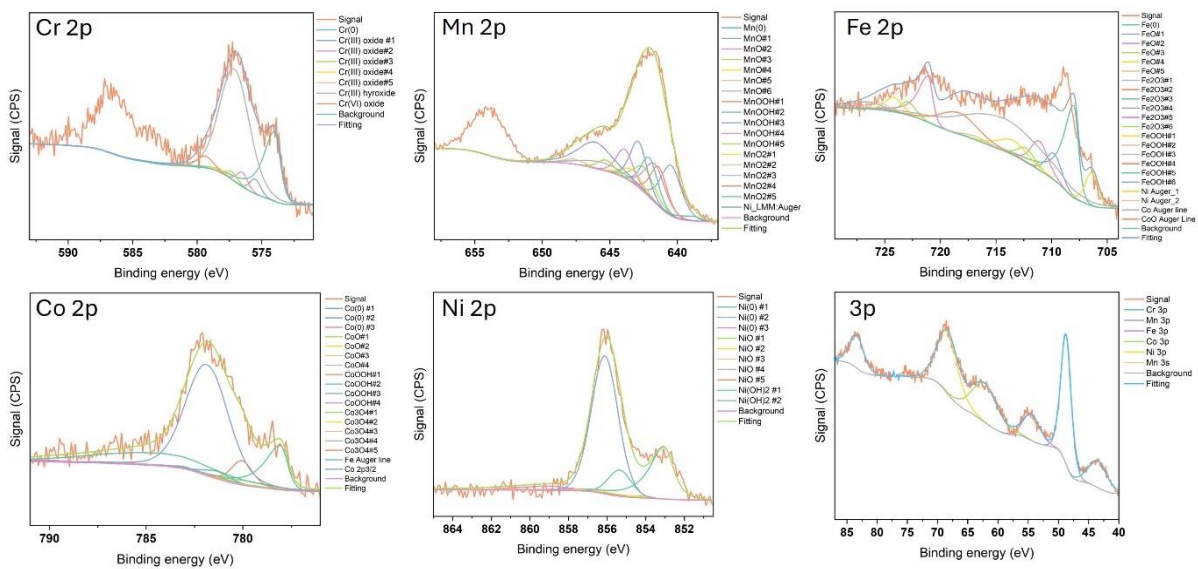

*Figure S25 Deconvoluted Cr 2p, Mn 2p, Fe 2p, Co 2p, Ni 2p and 3p XPS spectra of HEA NPs synthesized in acetonitrile within the size set 20nm < d < 10 nm.*

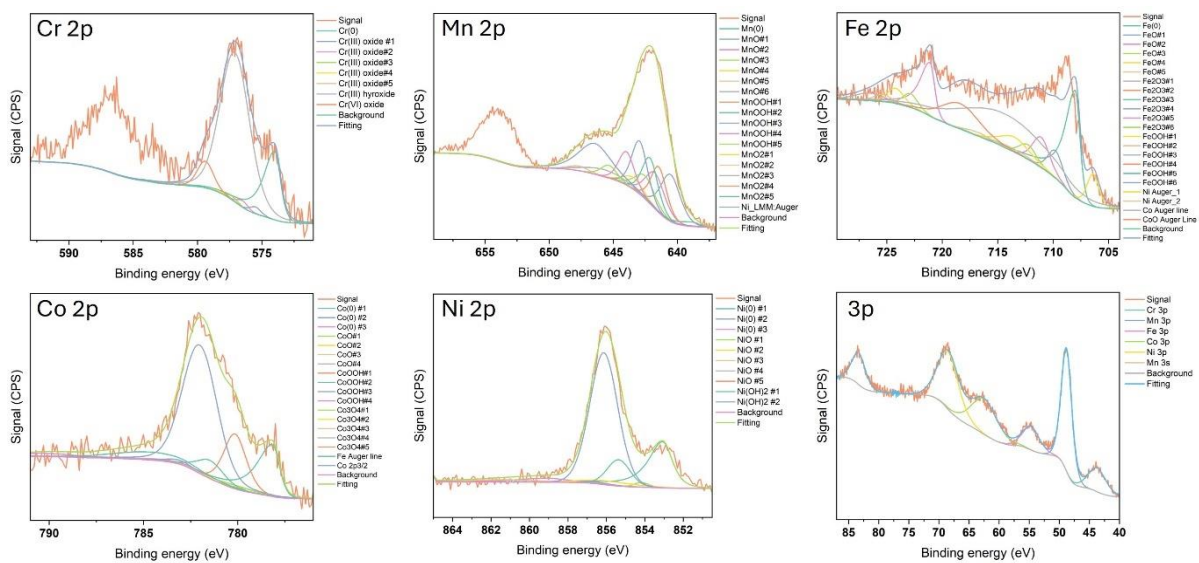

*Figure S26 Deconvoluted Cr 2p, Mn 2p, Fe 2p, Co 2p, Ni 2p and 3p XPS spectra of HEA NPs synthesized in acetonitrile within the size set d < 10 nm.*

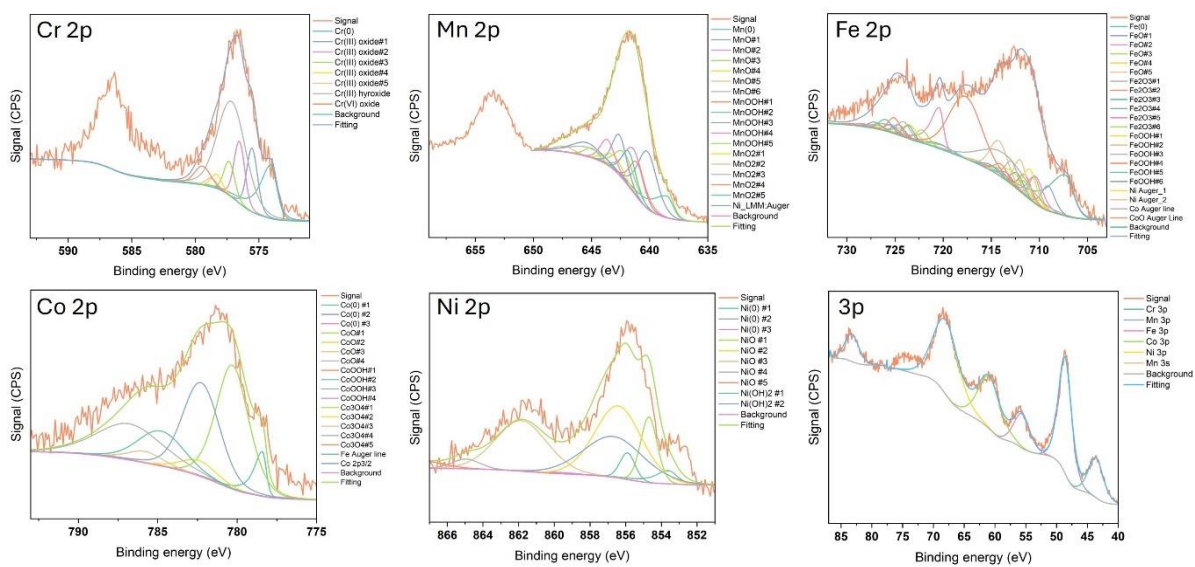

*Figure S27 Deconvoluted Cr 2p, Mn 2p, Fe 2p, Co 2p, Ni 2p and 3p XPS spectra of HEA NPs synthesized in acetone within the size set d > 100 nm.*

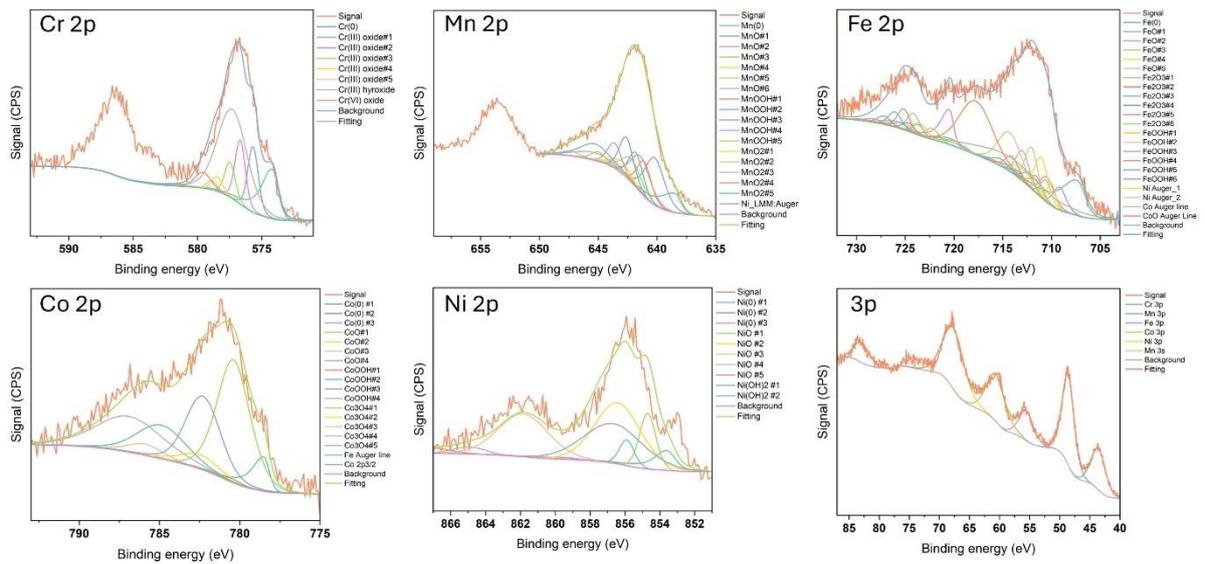

*Figure S28 Deconvoluted Cr 2p, Mn 2p, Fe 2p, Co 2p, Ni 2p and 3p XPS spectra of HEA NPs synthesized in ethanol within the size set d > 100 nm.*

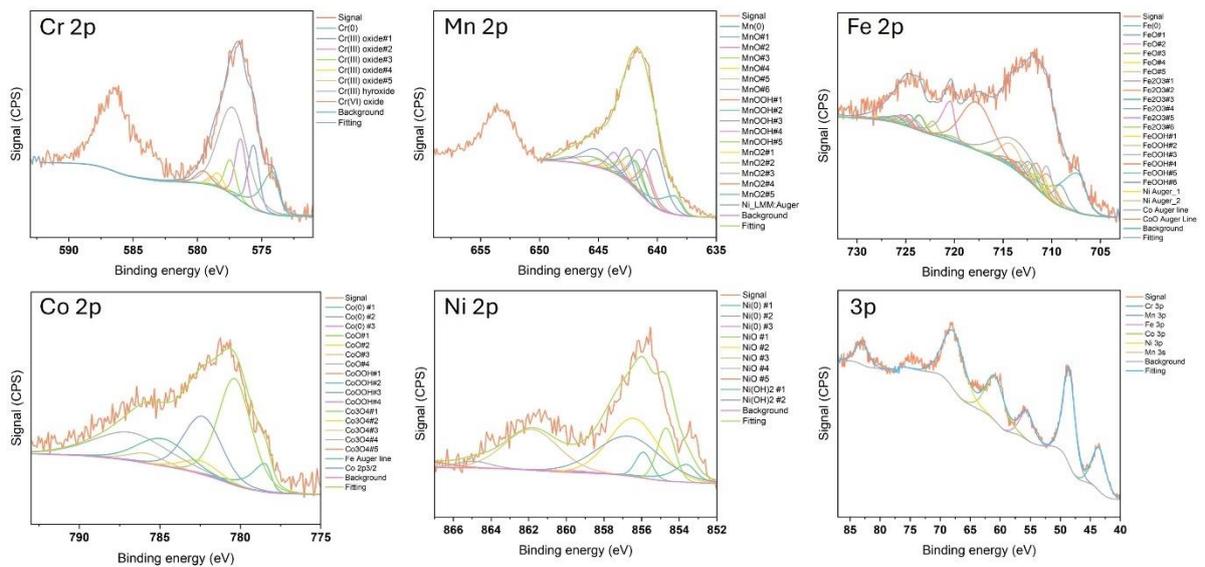

*Figure S29 Deconvoluted Cr 2p, Mn 2p, Fe 2p, Co 2p, Ni 2p and 3p XPS spectra of HEA NPs synthesized in ethanol within the size set 100 nm < d < 50 nm.*

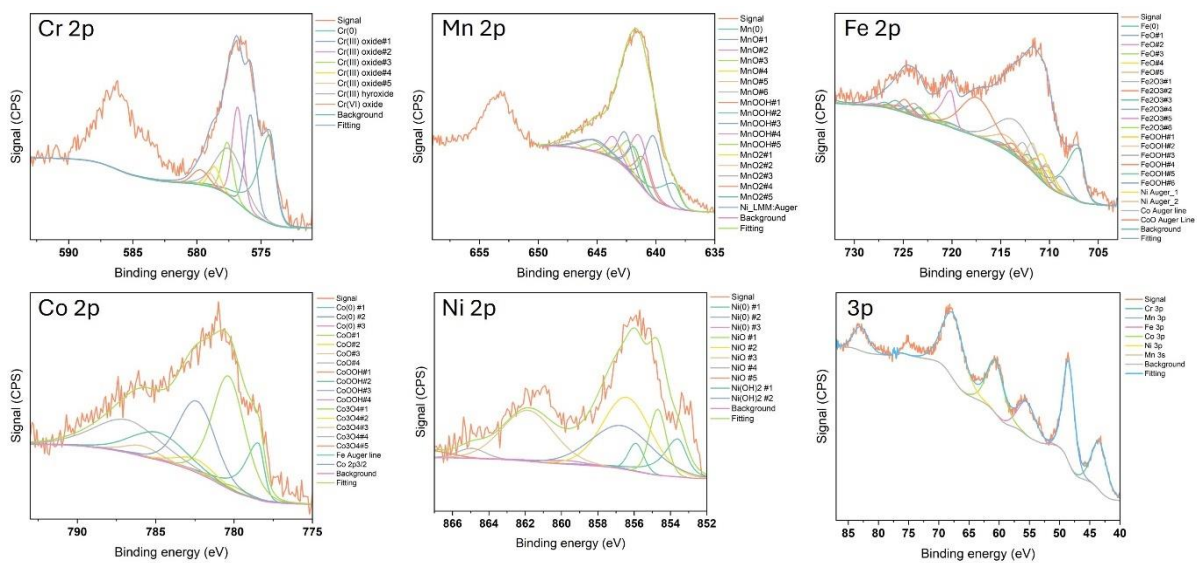

*Figure S30 Deconvoluted Cr 2p, Mn 2p, Fe 2p, Co 2p, Ni 2p and 3p XPS spectra of HEA NPs synthesized in ethanol within the size set 50 nm < d < 20 nm.*

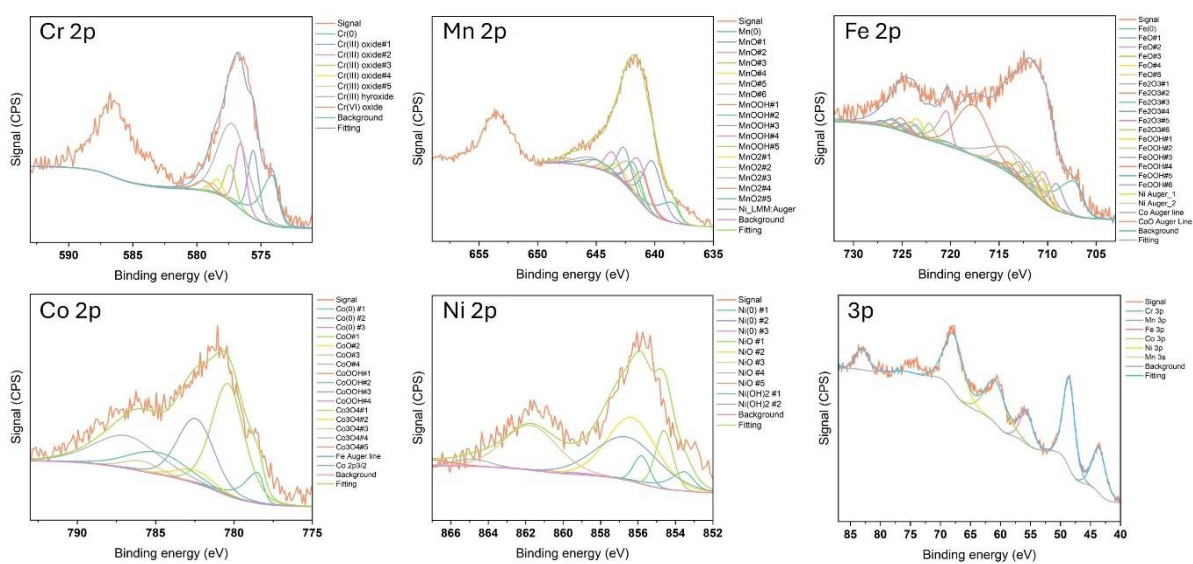

*Figure S31 Deconvoluted Cr 2p, Mn 2p, Fe 2p, Co 2p, Ni 2p and 3p XPS spectra of HEA NPs synthesized in ethanol within the size set 20 nm < d < 10 nm.*